\documentclass[12pt]{article}
\usepackage[a4paper,top=2.5cm,bottom=2.5cm,left=2.5cm,right=2.5cm,marginparwidth=1.75cm]{geometry}

\usepackage{setspace}
\usepackage{url} 
\usepackage{booktabs} 
\usepackage{times}
\usepackage{amsmath}
\usepackage{amsfonts} 
\usepackage{bm} 
\usepackage{float} 
\usepackage[round]{natbib}
\usepackage{amssymb} 
\usepackage{graphicx}
\usepackage{hhline}
\usepackage{parskip} 
\usepackage{xcolor} 

\title{Bayesian sample size determination using robust commensurate priors with interpretable discrepancy weights}

\author{Lou E. Whitehead\textsuperscript{1}*, \\
James M. S. Wason\textsuperscript{1},\\
Oliver Sailer\textsuperscript{2},\\
Haiyan Zheng \textsuperscript{3}\\
\small1. Biostatistics Research Group, Population Health Sciences Institute,\\
\small Newcastle University, Newcastle upon Tyne, UK\\
\small2. Boehringer Ingelheim  Pharma GmbH \& Co. KG, Biberach, Germany\\
\small3. Department of Mathematical Sciences, University of Bath, Bath, UK\\
\small*Corresponding Author Details: \\
\small Lou E. Whitehead; Address: Biostatistics Research Group, \\
\small Population Health Sciences Institute, \\
\small Ridley 1 Building, Richardson Road, Newcastle upon Tyne NE1 7RU, UK\\
\small Telephone number: +44(0)191-208-6000; E-mail address: l.whitehead2@newcastle.ac.uk}

\begin{document}
\bibliographystyle{apalike}
\date{}
\begin{singlespace}
\maketitle
\thispagestyle{empty}
\begin{abstract}
Randomized controlled clinical trials provide the gold standard for evidence generation in relation to the efficacy of a new treatment in medical research. Relevant information from previous studies may be desirable to incorporate in the design and analysis of a new trial, with the Bayesian paradigm providing a coherent framework to formally incorporate prior knowledge. Many established methods involve the use of a discounting factor, sometimes related to a measure of `similarity' between historical and the new trials. However, it is often the case that the sample size is highly nonlinear in those discounting factors. This hinders communication with subject-matter experts to elicit sensible values for borrowing strength at the trial design stage. Focusing on a commensurate predictive prior method that can incorporate historical data from multiple sources, we highlight a particular issue of nonmonotonicity and explain why this causes issues with interpretability of the discounting factors (hereafter referred to as `weights'). We propose a solution for this, from which an analytical sample size formula is derived. We then propose a linearization technique such that the sample size changes uniformly over the weights. Our approach leads to interpretable weights that represent the probability that historical data are (ir)relevant to the new trial, and could therefore facilitate easier elicitation of expert opinion on their values.

\end{abstract}

\maketitle
\end{singlespace}
\clearpage
\pagenumbering{arabic}
\section{Introduction}
\label{s:intro}
In clinical drug development randomized controlled trials (RCTs) are regarded as the gold standard for evaluating the efficacy of new treatments or interventions. Randomization of trial participants to the new treatment or a control group aims to reduce bias and provide a rigorous tool to examine whether a causal relationship exists between an intervention and outcome \citep{Hariton2018}. Sample size calculations are an essential part of clinical trial design, with a sample needing to be at least large enough to meet the study objectives but also small enough to minimize (for example) ethical or cost concerns \citep{julious2023sample}. In the frequentist paradigm, the number of participants recruited onto a study is often chosen to control the type I error rate (the rate of incorrectly declaring a treatment efficacious) and power (the rate of correctly declaring a treatment efficacious) to pre-specified levels, based on assumptions about the sampling distribution of the data and the size of the treatment effect considered clinically meaningful. 

Designing a trial with a large enough sample size to achieve the frequentist power can sometimes be infeasible, especially when there are limited numbers of participants available. 
This might be the case, for example, in rare disease trials or trials in pediatric populations. Pre-trial information, from historical studies conducted under similar circumstances, or elicited directly from expert opinion, could be useful to overcome this challenge, with the Bayesian paradigm offering a powerful tool to formalize this approach. In the Bayesian framework, a prior distribution is formed for a parameter of interest, which is then updated by the observed data to form a posterior distribution from which inferences can be made. Instead of designing a trial around frequentist type I error rates and power, Bayesian designs rely on alternative metrics for success; for instance, specification of posterior decision thresholds (the level of confidence we desire to have that a treatment is efficacious or futile), or the width or coverage probabilities of Bayesian credible intervals. The application of Bayesian methodology for trial design to the specific areas noted above has been considered in the literature, for example, by \citet{Hampson2014} for trials in very rare diseases, and \citet{Wadsworth2018} for pediatric studies.

  \citet{NeuenschwanderMAP2010} classify Bayesian methods for clinical trial design incorporating historical data according to the approach of constructing a prior distribution for a parameter of interest as follows:
\begin{itemize}
    \item `Irrelevance', where a prior is formed without reference to previous studies.
    \item `Similar', also termed `exchangeable', where a prior is formed by assuming that the parameter of interest in the new trial has been generated from the same underlying distribution as the parameter(s) in the historical trial(s). The meta-analytic predictive (MAP) prior proposed in \citet{NeuenschwanderMAP2010} is based on this assumption, with the authors noting the importance of careful selection of relevant historical data to render the exchangeability assumption plausible. A robust extension \citep{SchmidliRobustMAP2014} aims to effectively discount historical data in the case of prior/data conflict by using a weighted mixture distribution consisting of the MAP prior and a weakly informative component. 
    \item `Equal but discounted', which assumes parameters are the same, but discounts the precision of the parameter in the historical trial(s). The `power prior' suggested by \citet{ibrahim2000power} takes this approach, whereby historical evidence is downweighted by taking its likelihood to a power, $a \, \epsilon \, [0, 1]$.
    \item `Biased', which assumes historical parameters are potentially biased versions of the parameter in the new trial. The `commensurate prior' \citep{hobbs2011hierarchical, Hobbs2012CommensPrior} comes under this category, where historical information is downweighted by a commensurability parameter to form a predictive prior for the new study. The commensurability parameter directly parameterizes the similarity between each historical source and new data.
    \item ‘Equal’, equivalent to pooling historical data with the new study data.
\end{itemize}
The importance of carefully selecting historical trials to be included for planning a new trial is well understood. If the assumption of similarity is not satisfied, this can result in increased mean square error (MSE) of point estimates due to bias and either reduced power or increased type I error rate depending on the direction of the bias \citep{Viele2014}. Conversely, incorporation of quality historical information allows for reduced MSE and increased power (or reduced type I error rate) within the new trial.  A seminal paper by \citet{POCOCK1976175} provided a set of criteria for assessing the comparability between historical and current trials. Expert elicitation can play an important role in assessing comparability and helping to choose model parameters but the elicitation process is not trivial \citep{dias2017elicitation}. \citet{johnson2010methods} review different methods to elicit beliefs for Bayesian priors. 

This paper focuses on the design of a new two-arm RCT incorporating historical data from similar RCTs. We follow the series of research in sample size determination based on `commensurate priors' in \citet{Zheng2023a} in which the use of discrepancy weights $\epsilon \, [0, 1]$ quantifying the probability of (ir)relevance of information from multiple historical sources (with respect to the new trial) was recommended. The methodology in \citet{Zheng2023a} was later extended to basket trials in \citet{Zheng2023b}. In the setting of borrowing from historical data, specification of study-specific discrepancy weights at the design stage provides an explicit opportunity to make judgments concerning the relevance and rigor of past studies with respect to the new study \citep{NeuenschwanderMAP2010}. Furthermore, the elicitation of study-specific discrepancy weights may be more intuitive than eliciting model parameters of a distribution. 

It is the intention that the discrepancy weights recommended in \citet{Zheng2023a} act uniformly with respect to the amount of information that would subsequently be incorporated from a particular source. For example, specifying a historical study-specific weight of $0.50$ should result in incorporation of $50\%$ of the information from that source into the new trial design. In Section \ref{s:problemformulation} we demonstrate that this is not the case, and the weights in fact exhibit undesirable highly nonlinear behaviour. Of primary concern is nonmonotonicity, caused by the method used to aggregate information from multiple sources into a single prior, which hinders interpretability and makes elicitation of such weights difficult. Additional nonlinearity is also an issue, whereby small values of weights result in faster changes in the amount of information incorporated into the prior than their complement. We propose a solution in two parts. Firstly, in Section \ref{s:methods}, an alternative method of prior aggregation is proposed, for which the nonlinearity then has a simpler pattern, and from which a Bayesian sample size formula is derived. Secondly, a technique for linearization is provided such that the weights provide uniform shrinkage with respect to the sample size. The aim is to make interpretability simpler and thereby facilitate easier elicitation of such values. Section \ref{s:motivatingexample} provides a motivating example in which a sample size is sought for a hypothetical new RCT using historical data from several real-life historical clinical trials. Section \ref{s:performanceevaluation} presents a brief simulation study confirming pre-specified statistical properties are preserved across a range of scenarios with sample sizes determined according to our method. We finish with a discussion highlighting areas for future research in Section \ref{s:discussion}. 
\section{Problem Formulation}
\label{s:problemformulation}
Consider planning a two-arm randomized controlled superiority trial (referred to as `new trial' in the following) to evaluate an investigational treatment or intervention. Let $Y_{ij}$ be the measured post-randomization outcomes in the new trial for patient $i = 1, ..., n_j$ in treatment group $j = T, C$. Explicitly, $j=T$ refers to the experimental treatment group and $j=C$ refers to the control group. We assume outcomes are normally distributed with common variance in the outcome measures such that $Y_{ij} \sim \mathcal{N} (\mu_j, \sigma_0^2)$.
The groupwise sample means therefore follow a normal distribution, $\Bar{Y}_{j} \sim \mathcal{N} (\mu_j, \frac{\sigma_0^2}{n_j})$. Considering the distribution of the difference in group means leads to
\begin{equation*}
   \Bar{Y}_{T} - \Bar{Y}_{C} = \Bar{Y}_{\Delta}  \sim \mathcal{N} \left(\mu_{\Delta}, \frac{\sigma_0^2}{n R (1-R)}\right),
\end{equation*}
where the parameter $\mu_\Delta = \mu_T - \mu_C$  is our primary inferential target. $n = \sum_{j=T, C} n_j $ are the total number of trial participants randomized (to treatment or control) at the initiation of the trial and $R = n_T/n$ is the proportion randomly assigned to the experimental treatment arm.

In the Bayesian framework with no borrowing from historical data (for assumed known $\sigma_0^2$), a prior for $\mu_\Delta$ is specified,
\begin{equation*}
\label{eq:prior_nb}
    \mu_{\Delta} \sim \mathcal{N} \left(\mu_0, s_0^2\right),
\end{equation*}
which is then updated by the trial data 
to give a posterior distribution,
\begin{equation*}
\label{eq:nb_post_dist}
    \mu_{\Delta}| \bm{y_{new}} \sim \mathcal{N} \left(d_{\theta_0}, \sigma^2_{\theta_0}\right).
\end{equation*}
The posterior mean is given by
\begin{equation}
\label{eq:d_theta_nb}
    d_{\theta_0} = \frac{ \mu_0 \cdot s_0^{-2}  + (\Bar{y}_T - \Bar{y}_C) \cdot nR(1-R)/\sigma_0^2} {s_0^{-2}+nR(1-R)/\sigma_0^2},
\end{equation}
and the posterior variance is 
\begin{equation}
\label{eq:sig_nb}
\sigma^2_{\theta_0} = \left( \frac{1}{s_0^2}+ \frac{n R(1-R)}{\sigma_0^2} \right) ^{-1}.
\end{equation}
\subsection{Formulating Priors from Multiple Historical Sources}
\label{qpriors}
Suppose instead that there are $Q$ sources of historical data, $\bm{y_1}, ..., \bm{y_Q}$, that are relevant to incorporate in the planning of the new trial. 
$\lambda_q$ are the parameter counterparts of $\mu_{\Delta}$ in the historical trials and it is assumed they have been summarized by posterior distributions, $ \lambda_q\sim \mathcal{N} (\theta_{q}, \tau_q^2)$. Defining $\mu_{\Delta(q)}$ as the prediction for $\mu_{\Delta}$ in the new trial based on the information from trial $q$ alone, a set of $Q$ commensurate predictive prior distributions for $\mu_{\Delta}$ are formed centered on each $\theta_{q}$,
\begin{gather}
\label{eq:predprior}
\mu_{\Delta(1)} \sim \mathcal{N} (\theta_1, \xi_1^2) ,\ \ ... ,\ \ \mu_{\Delta(Q)} \sim \mathcal{N} (\theta_Q, \xi_Q^2).    
\end{gather}
We let $\xi_q^2 = \tau^2_q+\nu_{q}^{-1}$, where $\nu_{q}$ parameterizes the `commensurability' \citep{Zheng2023b} between $\lambda_q$ and $\mu_{\Delta}$ in terms of precision (further details are given in the following section).
\subsection{Estimating $\xi^2_{q}$}
\label{prio_for_nu} 
To quantify the relevance of each historical data source in respect of the new experiment, Zheng et al. introduce discrepancy parameters,
$\bm{w_q} = \{w_1, ...., w_Q\}$. These are prior weights $ \in [0, 1]$ intended to represent preliminary skepticism about how similar $\lambda_{q}$ and $\mu_{\Delta}$ are \citep{Zheng2023a}. These weights are incorporated into a Gamma mixture prior for the precision parameter, $\nu_{q}$:
\begin{gather}
\label{eq:e_prior}
    \nu_{q} \sim w_q \: Ga(a_{01}, b_{01}) + (1 - w_q) \: Ga(a_{02}, b_{02}),
\end{gather}
with $w_{q} \, \epsilon \,  [0,1], \:\:  {a_{01}, a_{02} > 1}$. This mixture prior is favoured for robust inferences as it offers flexible downweighting or borrowing from source $q$ depending on the value of $w_q$. Briefly, the values of $a_{01}, b_{01}$ are chosen such that the first Gamma mixture component has its mass on small values, therefore when $w_q \to 1$,  data from source $q$ is increasingly discounted. At the extreme, setting $w_q = 1$ indicates complete irrelevance of information from source $q$ to the new trial. On the other hand, values of $a_{02}, b_{02}$ are chosen such that the second Gamma mixture component has its mass on large values. In this case, setting $w_q \to 0$ results in a greater degree of incorporation of information from source $q$. Setting $w_q = 0$ indicates exchangeability between $\lambda_{q}$ and $\mu_{\Delta}$, i.e., $\mu_{\Delta(q)} \sim \mathcal{N}(\theta_q, \tau_q^2)$. It is anticipated in a real application that, at the design stage of a new trial, $\bm{w_q} \, \epsilon \, [0, 1]$ are chosen in collaboration with a subject-matter expert(s)  to reflect the anticipated degree of (ir)relevance between each historical trial and the new experiment.
As detailed in \citet{Zheng2023a}, the Gamma mixture prior in (\ref{eq:e_prior}) can be approximated by matching the first two moments of a unimodal \textit{t} mixture distribution. This leads to an approximation of the between-trial variance (i.e., between source $q$ and the new experiment),
\begin{gather}
\label{eq:nuint}
    \nu^{-1}_{q} \simeq \frac{w_{q} b_{01}}{a_{01}-1}+\frac{(1-w_{q}) b_{02}}{a_{02}-1}. \nonumber
\end{gather}
We note that if we were being fully Bayesian we would keep the prior for $\nu_q$ in its distributional form, however in this paper we are looking to propose an asymptotically approximate sample size formula and so we make a simplifying assumption. The variance between each source $q$ and the new trial is therefore estimated as   
\begin{gather*}
\label{eq:xi2_basedonnew_nu}
    \xi_q^2 = \tau_q^2 + \frac{w_{q} b_{01}}{a_{01}-1}+\frac{(1-w_{q}) b_{02}}{a_{02}-1}.
\end{gather*}
\subsection{Aggregating Multiple Distributions to Form a Collective Prior}
\label{CPP_Z}
In \citet{Zheng2023a}, an informative collective prior (hereafter referred to as `CP') is formed by aggregating the $Q$ predictive distributions in (\ref{eq:predprior}) into a single prior such that $\mu_{\Delta}| \bm{y_1},...,\bm{y_Q} \sim \mathcal{N}(\theta_{CP},  \sigma^2_{CP})$ using the convolution operator for the sum of normal random variables \citep{grinstead1997introduction}, where
\begin{gather*}
\label{eq:cpp_mu_sigma}
    \theta_{CP} = \sum_{q=1}^{Q} p_{q}\theta_{q}, \ \
    \sigma^2_{CP} = \sum_{q=1}^{Q} p^2_{q}\xi^2_{q}. 
\end{gather*}
$p_q$ are synthesis weights, set to a decreasing function of $w_q$, such that are all between $0$ and $1$ and sum to $1$. In \citet{Zheng2023a},
\begin{gather}
\label{eq:pqfunc}
        p_q = \frac{\exp(-w_q^2/c_0)}{\sum_{q=1}^{Q} \exp(-w_q^2/c_0)}, 
\end{gather}
where $c_0$ is a pre-defined concentration parameter which governs how much influence $w_q$ have on $p_q$. Further details on the $p_q$ function in (\ref{eq:pqfunc}) and how to choose $c_0$ are provided in \citet{ZhengandWason2022} and \citet{Zheng2023a}. 
The CP is updated by the trial data to give the posterior,
\begin{equation*}
\label{eq:post_dist}
    \mu_{\Delta}| \bm{y_1},...,\bm{y_Q}, \bm{y_{new}} \sim \mathcal{N}\left(d_{\theta_1}, \sigma^2_{\theta_1} \right),
\end{equation*}
In the same way as Equations (\ref{eq:d_theta_nb}) and (\ref{eq:sig_nb}), the posterior mean and variance are given by 
\begin{equation*}
\label{eq:d_thetaCPP}
    d_{\theta_1} = \frac{\theta_{CP} \cdot \sigma^{-2}_{CP} + (\Bar{y}_T - \Bar{y}_C) \cdot nR(1-R)/\sigma_0^2} {\sigma^{-2}_{CP}+nR(1-R)/\sigma_0^2}
\end{equation*}
and 
\begin{equation*}
\label{eq:precthetaCPP}
 \sigma^2_{\theta_1}=\left(\sigma^{-2}_{CP}+ \frac{n R(1-R)}{\sigma_0^2}\right)^{-1}.   
\end{equation*}
\subsection{Varying $w_q$ to Alter the Amount of Information from Source $q$}
\label{varyingw}
\begin{figure}[H]
    \centering
    \includegraphics[scale=0.35]{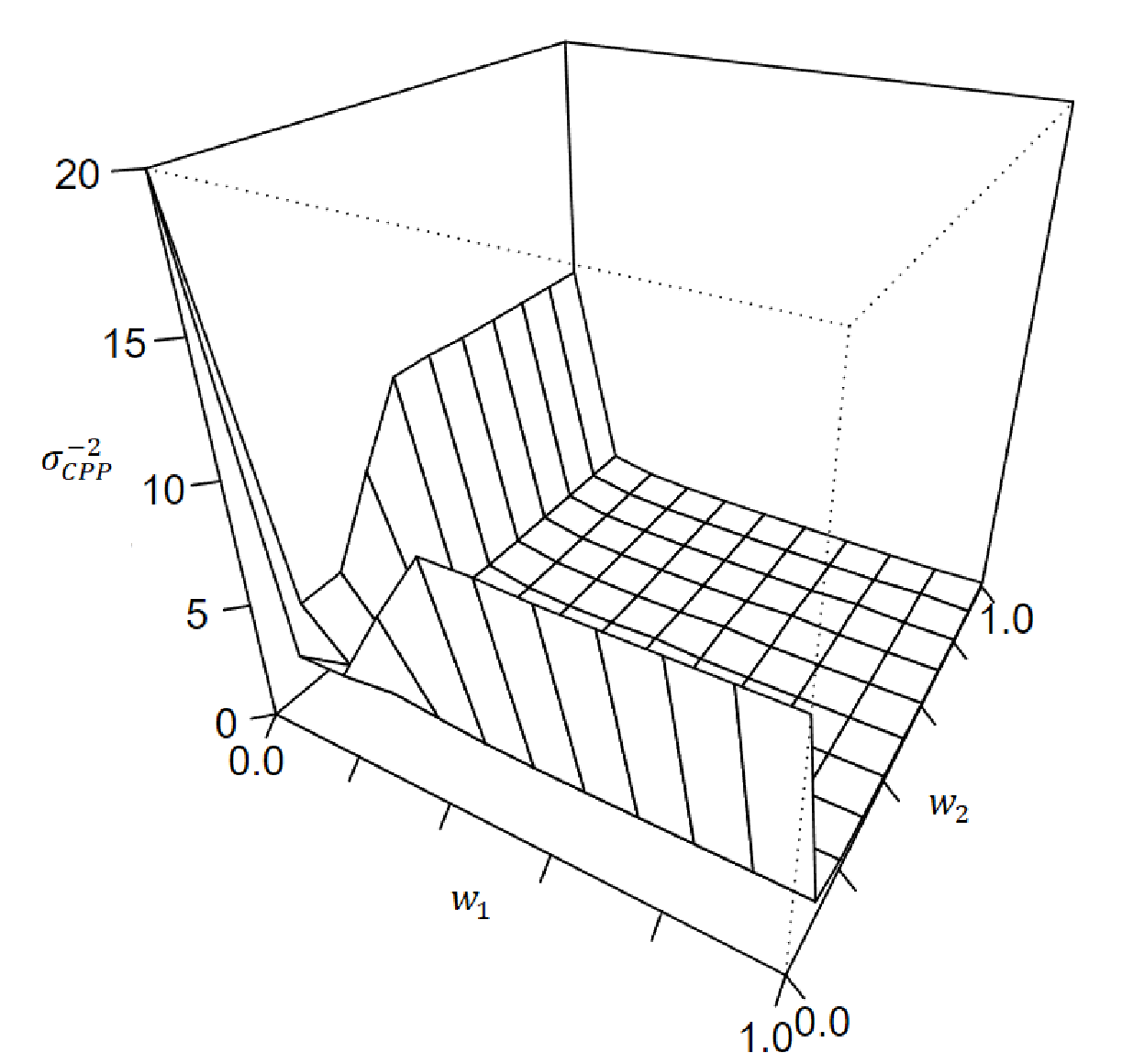}
    \caption{CP precision, $\sigma_{CP}^{-2}$ (Equation \ref{eq:precfullCPP}), with respect to varying discrepancy weights, $w_1$ and $w_2$, for borrowing from two historical sources, $q=1,2$. Undesirable nonmonotonicity can clearly be seen around $w_1 = 0$ and $w_2 = 0$.}
    \label{fig:Precision_CPP1_w1w2}
\end{figure}

For fixed $\bm{\tau^2_q}, a_{01}, b_{01}, a_{02}, b_{02}$, $c_0$, the CP precision $\sigma_{CP}^{-2}$ is a function of $\bm{w_q}$ and is a measure of the amount of prior information on the treatment effect in the new trial (which varies depending on the values of $\bm{w_q}$),
\begin{align}
\label{eq:precfullCPP}
    \sigma^{-2}_{CP} & = \left[\sum_{q=1}^{Q} p^2_{q} \xi^{2}_{q}\right]^{-1} \nonumber \\
    & = \left[\sum_{q=1}^{Q} p^2_{q}\left(\tau_q^2 + \frac{w_{q} b_{01}}{a_{01}-1}+\frac{(1-w_{q}) b_{02}}{a_{02}-1}\right) \right]^{-1}.
\end{align}
In Figure \ref{fig:Precision_CPP1_w1w2}, we visualize how $\sigma_{CP}^{-2}$ varies according to $\bm{w_q}$ in an example when $Q=2$.
For illustrative purposes, values of all other parameters are held fixed ($\tau_1^2=\tau_2^2=0.1, a_{01}=1.1, b_{01}=1.1, a_{02}=1 \times 10^6, b_{02}=1, c_0= 0.05$). It can be seen that for $w_1=w_2=0$, corresponding to full incorporation of information from both historical sources, the CP precision is maximized (as desired). Similarly, for $w_1=w_2=1$, corresponding to full discounting of information from both sources, the CP precision is minimized (as desired). 
We nonetheless also see the (undesirable) highly nonlinear nature of $\sigma_{CP}^{-2}$ with respect to varying $\bm{w_q}$ in two respects. Firstly, it is clear that the majority of the change in prior precision occurs for small values of $w_q$; beyond around $w_q > 0.2$, there is almost no discernible change in $\sigma_{CP}^{-2}$. Assuming that $\bm{w_q}$ are expert elicited probabilities, this could result in a large loss of information because specifying any $\bm{w_q} \, \gtrapprox \, 0.2$ will result in almost full discounting of data from source $q$. This issue occurs in varying degrees for any value of $Q$ and regardless of the values that the other parameters are fixed at. 

Secondly, and more importantly, when $Q>1$, local minima/maxima can be seen around $w_1=0$ and $w_2=0$. This nonmonotonic behaviour in Equation (\ref{eq:precfullCPP}) occurs whenever $Q>1$ due the method of prior aggregation as well as the higher order terms in $w_q$ introduced by the synthesis weighting function, Equation (\ref{eq:pqfunc}). This is in contrast to how we would fundamentally wish the discrepancy weights to behave; it should be the case that increasing $\bm{w_q}$ always leads to decreasing $\sigma^{-2}_{CP}$.

These two issues mean that $\bm{w_q}$ are not interpretable as probabilities and hinder communication with subject-matter experts to elicit sensible values at the trial design stage. 

An alternative method of prior aggregation (and therefore a new way of formulating the CP precision) is necessary so that the nonlinearity has a simpler form. Specifically, the CP precision should be monotonically decreasing with respect to increasing $\bm{w_q}$. Details of our proposal are given in Section \ref{CPP_W}. Following derivation of a Bayesian sample size formula in Section \ref{Bayes_SS}, we also seek to recalibrate the weights. This is achieved in Section \ref{w_trans_process} via a functional transformation of each $w_q \to w'_q$, where $w'_q = f(w_q)$, such that the prior precision (and therefore the derived sample size function) varies linearly with respect to $w_q \, \epsilon \,  [0, 1]$.
 
\section{Methods}
\label{s:methods}
\subsection{Proposed Method of Prior Aggregation}
\label{CPP_W}
Following the set of predictive priors in (\ref{eq:predprior}), we propose an alternative method of prior aggregation suggested in \citet{winkler1981combining}. This results in a new collective prior, $\mu_{\Delta}| \bm{y_1},...,\bm{y_Q} \sim N (\theta_{CP^*},  \sigma^2_{CP^*})$, where
\begin{equation*}
\label{eq:cppWink_mu_sigma}
    \theta_{CP^*} = \sum_{q=1}^{Q} p^{*}_{q}\theta_{q}, \ \
    \sigma^2_{CP^*} = \left(\sum_{q=1}^{Q} \xi^{-2}_{q}\right)^{-1},
\end{equation*}
\begin{equation*}
    \label{eq:Winkler_weighting_function}
    p^{*}_q = \frac{\xi^{-2}_q} {\left(\sum_{q=1}^{Q} \xi^{-2}_{q}\right)}.
\end{equation*}

As in Section \ref{CPP_Z}, the CP mean, $\theta_{CP^*}$, is a weighted linear sum of the means from (\ref{eq:predprior}). Synthesis weights $p^{*}_{q}$ now incorporate information on both $\bm{\tau_q^2}$ and $\bm{w_q}$, rather than only $\bm{w_q}$ as in Equation (\ref{eq:pqfunc}) (since $\xi^{-2}_{q} = (\tau_q^2 + \frac{w_{q} b_{01}}{a_{01}-1}+\frac{(1-w_{q}) b_{02}}{a_{02}-1})^{-1}$). This preserves the desirable property that smaller $w_q$ correspond to larger $p^{*}_{q}$, and introduces the (also desirable) property that smaller $\tau_q^2$ correspond to larger $p^{*}_{q}$. As required, $p^{*}_q$ sum to $1$ and are all between $0$ and $1$.

The CP variance, $\sigma^2_{CP^*}$, is the reciprocal of the sum of the precisions, $\xi^{-2}_q$. Again, this preserves the desirable property that a smaller $w_q$ results in source $q$ receiving a larger weight in $\sigma^{-2}_{CP^*}$. The formulation of the CP mean and variance in this manner is exactly in line with the theory of Bayesian updating of normal distributions with conjugate priors, with an initial noninformative prior for $\mu_{\Delta}$ (as discussed in \citet{winkler1981combining}). 

\begin{figure}[H]
    \centering
    \includegraphics[scale=0.35]{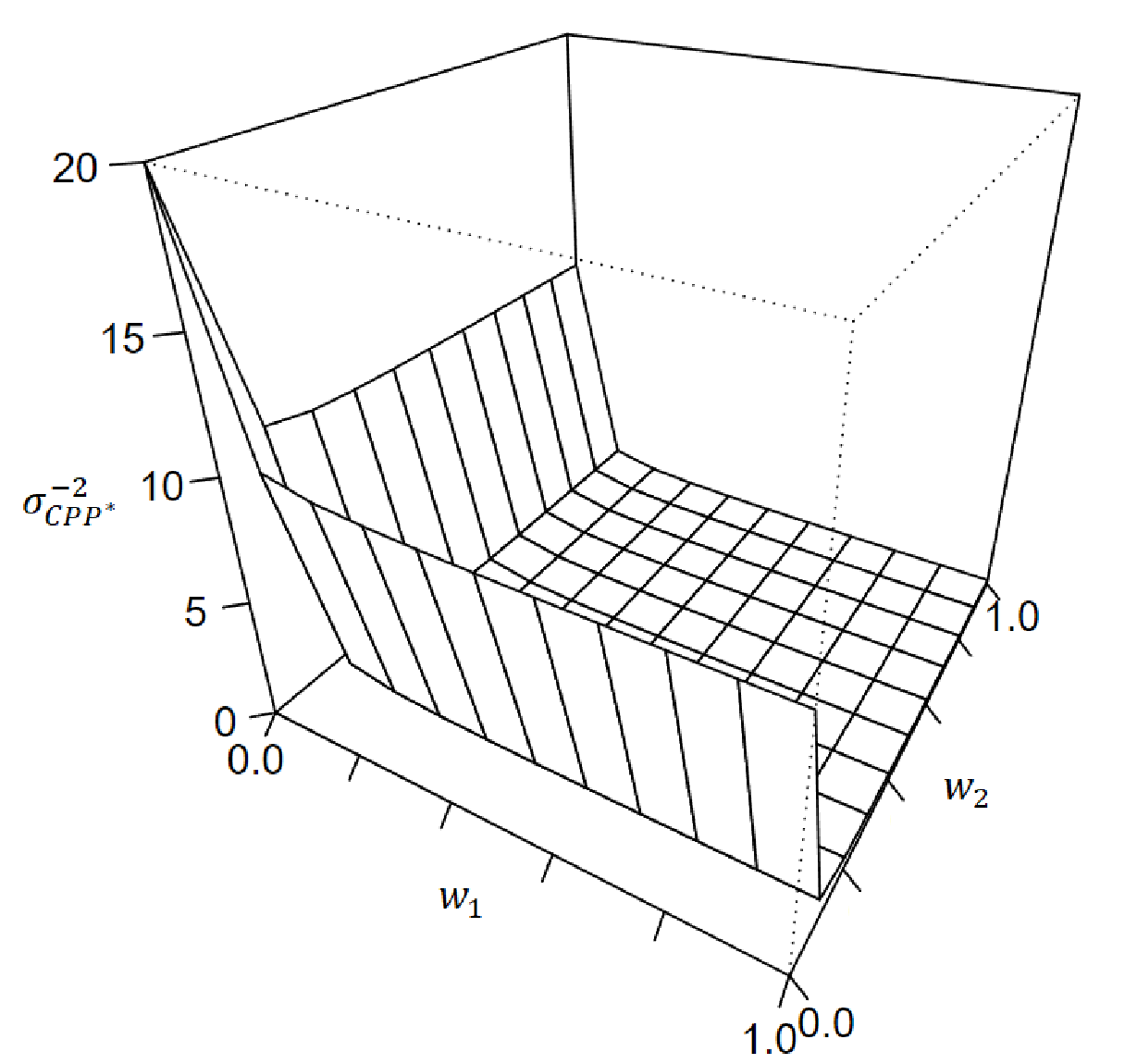}
    \caption{Proposed CP precision, $\sigma_{CP^*}^{-2}$ (Equation (\ref{eq:precnew})), with respect to varying discrepancy weights, $w_1$ and $w_2$, for borrowing from two historical sources, $q=1,2$. The prior precision is now monotonically decreasing with respect to increasing $\bm{w_q}$.}
    \label{fig:Precision_CPPnew_w1w2}
\end{figure}

 We note that an advantage of both the proposed prior aggregation method and the method detailed in Section \ref{CPP_Z} is that they allow for analytic sample size calculations. The proposed aggregation method preserves the desirable properties of the previous method of prior aggregation (described above) as well as fitting neatly into our Bayesian framework. 
However, central to the purpose of this paper, the nonlinearity of $\sigma^{-2}_{CP^*} $ in Equation (\ref{eq:precnew}) with respect to varying $\bm{w_q}$ now has a simpler pattern when compared with Equation (\ref{eq:precfullCPP}). Specifically, its first derivative with respect to each $w_q$ is always negative and therefore it is monotonically decreasing. In contrast to Equation (\ref{eq:precfullCPP}), this is now how we would wish $\bm{w_q}$ to behave - i.e., increasing $w_q$ should always lead to decreased prior precision. This is visualized in Figure \ref{fig:Precision_CPPnew_w1w2} using the same parameters as Figure \ref{fig:Precision_CPP1_w1w2} for borrowing from two historical datasets.
Additional motivation to use this method of aggregation in our particular application is that terms in the CP precision relating to each $w_q$ are now linearly independent of each other, i.e., 
\begin{align}
\label{eq:precnew}
    \sigma^{-2}_{CP^*} &= \sum_{q=1}^{Q} \xi^{-2}_{q} \nonumber \\
    &=  \sum_{q=1}^{Q} \left[\left(\tau_q^2 + \frac{w_{q} b_{01}}{a_{01}-1}+\frac{(1-w_{q}) b_{02}}{a_{02}-1}\right)^{-1}\right].
\end{align}
This means that we can achieve linearization of $w_q$ with respect to sample size (details in the following sections), which would be impossible to achieve by the previous aggregation method due to the issue of nonmonotonicity.
As in Section \ref{eq:nb_post_dist}, the CP is updated by the trial data to give the posterior,
\begin{equation*}
\label{eq:post_dist_new}
    \mu_{\Delta}| \bm{y_1},...,\bm{y_Q}, \bm{y_{new}} \sim \mathcal{N}\left(d_{\theta^*}, \sigma^2_{\theta^*} \right),
\end{equation*}
where  
\begin{equation}
\label{eq:d_thetaCPPnew}
    d_{\theta^*} = \frac{\theta_{CP^*} \cdot \sigma^{-2}_{CP^*}   + (\Bar{y}_T - \Bar{y}_C) \cdot nR(1-R)/\sigma_0^2} {\sigma^{-2}_{CP^*}+nR(1-R)/\sigma_0^2}
\end{equation}
and  
\begin{equation}
\label{eq:precthetaCPPnew}
 \sigma^2_{\theta^*}=\left(\sigma^{-2}_{CP^*}+ \frac{n R(1-R)}{\sigma_0^2}\right)^{-1}.   
\end{equation}
\subsection{Bayesian Decision Making Framework}
\label{Bayesian_DM}
We now introduce a Bayesian decision making framework proposed in \citet{Whiteheadetal2008}. 
For pre-specified posterior decision thresholds $\eta$ and $\zeta$, we seek a sample size to guarantee we have sufficient evidence to conclude either efficacy or futility respectively. These thresholds represent the degree of evidence we would require to be convinced of efficacy or futility of treatment over control. Explicitly, if $\mathbb{P}(\mu_{\Delta}>0) > \eta$ then we conclude that the treatment is efficacious and if $\mathbb{P}(\mu_{\Delta} \leq \delta) > \zeta$ then we conclude that the treatment is futile, where $\eta$ and $\zeta \, \epsilon \, [0, 1]$ and $\delta$ is some minimally clinically important treatment effect size.

For a generic posterior distribution $\mu_{\Delta} \sim \mathcal{N}(d_{\theta}, \sigma^2_{\theta})$, the probability that the treatment effect is greater than zero is
\begin{equation*}
\label{eq:P_mu_gt_zero}
   \mathbb{P}(\mu_{\Delta}>0) = 1 - \Phi\left(-\frac{d_{\theta}}{\sigma_{\theta}} \right) = \Phi\left(\frac{d_{\theta}}{\sigma_{\theta}} \right), 
\end{equation*}
where $\Phi(\cdot)$ denotes the standard normal cumulative distribution function. Therefore, we will conclude convincing evidence of treatment benefit when $\frac{d_{\theta}}{\sigma_{\theta}} \geq z_{\eta}$, where $z_{\eta}$ satisfies $\Phi(z_{\eta})=\eta$. 

Similarly, the posterior probability that the treatment effect is less than (or equal to) $\delta$ is
\begin{equation*}
\label{eq:P_mu_lt_delta}
    \mathbb{P} (\mu_{\Delta} \leq \delta) = \Phi\left(\frac{\delta-d_{\theta}}{\sigma_{\theta}} \right).
\end{equation*}
Therefore, convincing evidence of treatment futility occurs when $\frac{\delta - d_{\theta}}{\sigma_{\theta}} \geq z_{\zeta}$, where $z_{\zeta}$ satisfies $\Phi(z_{\zeta})=\zeta$. 
\subsection{Bayesian Sample Size Formula}
\label{Bayes_SS}
Following the same approach detailed in \citet{Zheng2023b}, to reach a decisive conclusion regarding treatment efficacy, we require a large enough sample size such that either $d_{\theta}/\sigma_{\theta} \geq z_{\eta}$ or $(\delta - d_{\theta})/\sigma_{\theta} \geq z_{\zeta}$, i.e.,
\begin{equation*}
\label{eq:sumpowert1e}
    \frac{d_{\theta}}{\sigma_{\theta}} +\frac{(\delta - d_{\theta})}{\sigma_{\theta}} \geq z_{\eta} + z_{\zeta}.
\end{equation*}
Simplifying and rearranging, this is equivalent to requiring that
\begin{equation}
\label{eq:deriv1}
    \frac{1}{\sigma_{\theta}^2} \geq \left(\frac{z_{\eta}+z_{\zeta}}{\delta} \right)^2.
\end{equation}
We see that the left hand side of (\ref{eq:deriv1}) is equal to the posterior precision. Replacing $\sigma^2_{\theta}$ with the variance in (\ref{eq:sig_nb}), we therefore obtain a Bayesian sample size formula in the case of no borrowing,
\begin{equation}
\label{eq:ss_no_borrow}
    n \geq \frac{\sigma_{0}^2}{R(1-R)} \left( \left(\frac{z_{\eta}+z_{\zeta}}{\delta} \right)^2 - \frac{1}{s_{0}^2} \right).
\end{equation}
Note that if we wished to consider a purely frequentist formulation of the problem, then the necessary sample size is simply,
\begin{equation}
\label{eq:ss_freq}
    n \geq \frac{\sigma_{0}^2}{R(1-R)}  \left(\frac{z_{1-\alpha}+z_{1-\beta}}{\delta} \right)^2,
\end{equation}
where $\alpha$ and $\beta$ are the usual parameters set to control type I  and type II error rates respectively.

Replacing $\frac{1}{s_{0}^2}$ in (\ref{eq:ss_no_borrow}) with $\sigma_{CP^*}^{-2}$ from (\ref{eq:precnew}), we obtain our sample size calculation informed by $Q$ sources of historical data,
\begin{gather*}
\label{eq:ss_prec_Wink}
    n \geq \frac{\sigma_{0}^2}{R(1-R)} \left( \left(\frac{z_{\eta}+z_{\zeta}}{\delta} \right)^2 - \sigma_{CP^*}^{-2} \right),
\end{gather*} 
i.e.,
\begin{gather}
\label{eq:ss_with_borrow_Wink}
    n \geq \frac{\sigma_{0}^2}{R(1-R)} \left( \left(\frac{z_{\eta}+z_{\zeta}}{\delta} \right)^2 - \sum_{q=1}^{Q} \left[\left(\tau_q^2 + \frac{w_{q} b_{01}}{a_{01}-1}+\frac{(1-w_{q}) b_{02}}{a_{02}-1}\right)^{-1} \right] \right),
\end{gather}  
with $\bm{w_{q}} \, \epsilon \,  [0,1], \:\:  {a_{01}, a_{02} > 1}$. We note explicitly the assumptions embedded into this sample size formula, which are common to many normal models. The validity of the sample size calculation depends on these assumptions being satisfied:
\begin{enumerate}
\item Common (and known) variance in outcomes from the new trial.
\item Independence of observations.
\item Homoscedasticity and normality of residuals.
\end{enumerate}
For non-normal data, a suitably adapted formula based on the approach of constructing a normal test statistic in the Generalized Linear Model (GLM) framework via a transformation could be applied. In Supplementary Materials A.1, A2, and A.3,
we demonstrate this by deriving sample size formulas for RCTs with binary and time-to-event data, and for single-arm settings with binary outcomes.
\subsection{Interpretable Discrepancy Weights}
\label{w_trans_process}
We now detail the linearization steps which result in $\bm{w_q}$ that are directly interpretable as a degree of discrepancy on the information scale, $\epsilon \, [0, 1]$. The idea is similar to the idea of functional uniform priors proposed in \citet{Bornkamp2012, Bornkamp2014} for nonlinear regression, in which a method for formulating a prior for a parameter of interest is proposed such that that it is uniform in the space of functional shapes of the underlying nonlinear function. 
We start by isolating each nonlinear part of the sample size function in Equation (\ref{eq:ss_with_borrow_Wink}) with respect to $w_q$ (for fixed $\tau_q^2, a_{01}, b_{01}, a_{02}, b_{02}$). These are the individual precision terms making up $\sigma^{-2}_{CP^*}$ in Equation (\ref{eq:precnew}), i.e.,
\begin{equation}
\label{eq:nl_part_Qsource}
    \xi_q^{-2}(w_q) = \left(\tau_q^2 + \frac{w_{q} b_{01}}{a_{01}-1}+ \frac{(1-w_{q}) b_{02}}{a_{02}-1}\right)^{-1}.
\end{equation}
Step 1: Perform linear interpolation on (\ref{eq:nl_part_Qsource}):
\begin{equation}
\label{eq:f_lin_q}
    h(w_q)=(1-w_q)\xi_q^{-2}(0)+w_q \xi_q^{-2}(1) = \xi_q^{-2}(0)+w_q(\xi_q^{-2}(1)-\xi_q^{-2}(0)).
\end{equation}
This essentially `draws a line' between $\xi_q^{-2}(w_q = 0)$ and $\xi_q^{-2}(w_q = 1)$ so that changes in $\xi_q^{-2}$ (and therefore the corresponding sample size) are spread evenly across the full range of $w_q  \, \epsilon \,  [0,1]$. This also necessarily ensures that the mapping $w_q \to w'_q$ preserves the property that $w_q=0 \to w_q'=0$ and $w_q=1 \to w_q'=1$. 

Step 2: Find the inverse of (\ref{eq:nl_part_Qsource}). This allows calculation of any $w_q$ value corresponding to a given $\xi_q^{-2}$:
\begin{equation}
\label{eq:nl_part_rearr}
    g(\xi_q^{-2})=(\xi_q^{-2})^{(-1)}
\end{equation}
Step 3: Substitute the linearized $\xi_q^{-2}$ values obtained in (\ref{eq:f_lin_q}) into (\ref{eq:nl_part_rearr}):
\begin{equation}
\label{eq:simp_wdash_funcq}
     f(w_q) = g(h(w_q))= w'_q,
\end{equation}
$f(w_q) $ is now the necessary transformation of $w_q \to w'_q$. Now, if we obtain expert elicited values of $\bm{w_q}$, corresponding to a percentage degree of discrepancy between each historical source and the new trial, we can use Equation (\ref{eq:ss_with_borrow_Wink}) with $\bm{w'_q}$ to incorporate $((1-w_q) \times 100) \%$ of the information from the corresponding historical dataset in the planning of the new trial.

This transformation is possible due to the proposed method of prior aggregation. Unlike the original $\sigma^{-2}_{CP}$ in Equation (\ref{eq:precfullCPP}), the proposed $\sigma^{-2}_{CP^*}$ is a monotonic function in $w_q$, and each of the $\xi_q^{-2}$ terms (which form $\sigma^{-2}_{CP^*}$) are linearly independent (i.e., separated by the addition operator). The transformation procedure can easily be extended to any number of sources, $q=1,...,Q$, with each functional transformation of $w_q \to f(w_q)=w_q'$ being performed independently with no additional complexity.

\begin{figure}[H]
    \includegraphics[scale=0.75]{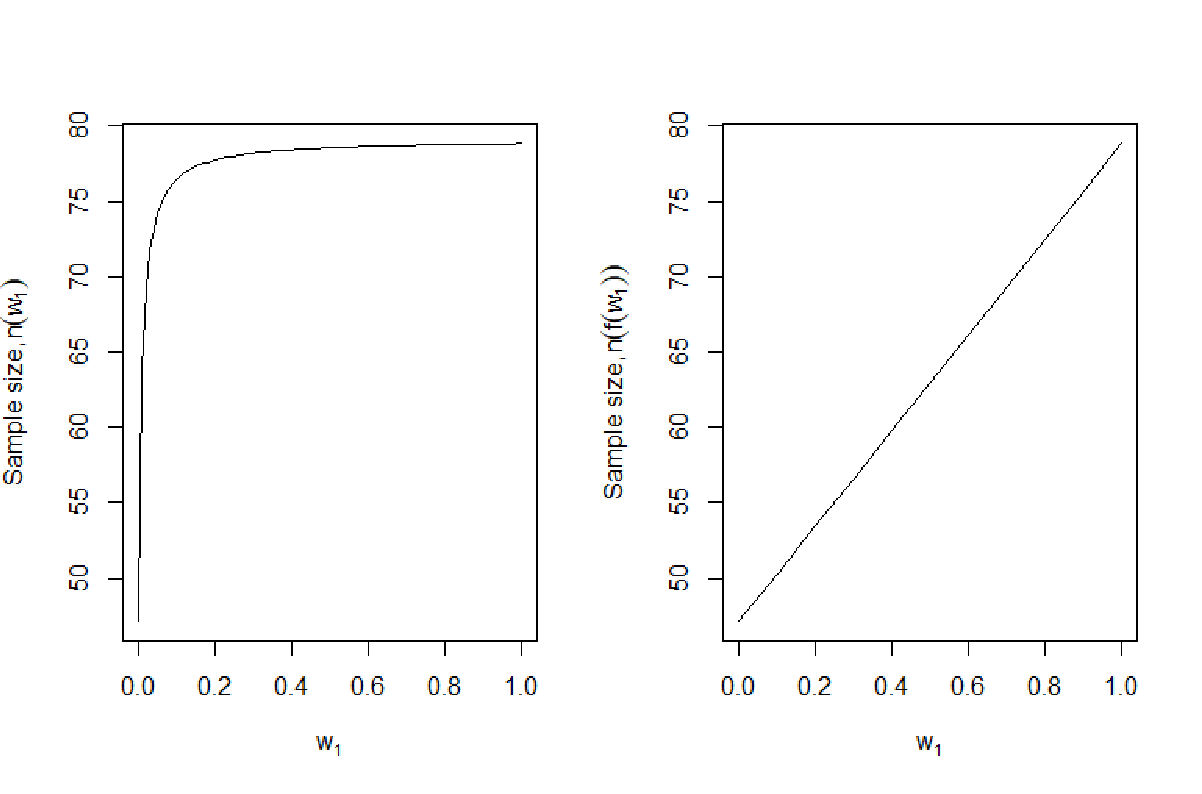}
    \caption{Sample size, $n$, with respect to varying $w_1$, for borrowing from a single source of data, both before (left) and after (right) functional transformation of $w_1 \to f(w_1)=w'_1$. Note that as required, minimum and maximum sample sizes corresponding to $w_1=0$ and $w_1=1$ respectively remain identical in both cases.}
    \label{fig:beforeandafter1}
\end{figure}

\begin{figure}[H]
    \includegraphics[scale=0.55]{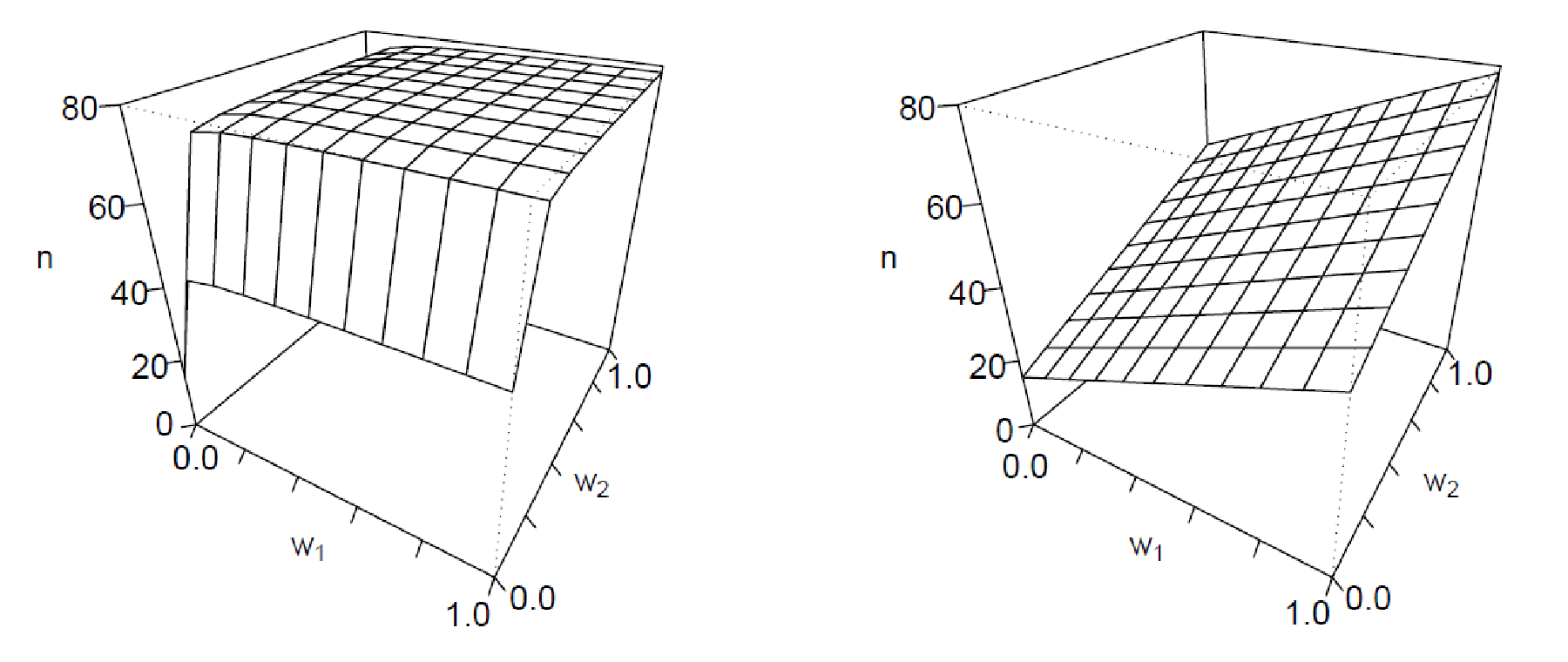}
    \caption{Sample size $n$ (vertical axis) borrowing from two historical sources, with respect to varying $w_1$ and  $w_2$. The left figure is before the functional transformation of $\bm{w_q} \to \bm{w'_q}$, i.e., $n(\bm{w_q}, \bm{\tau^2_q})$, the right figure is after, i.e., $n(\bm{w'_q}, \bm{\tau^2_q})$.}
    \label{fig:beforeandafter2}    
\end{figure}

The effect is visualized in Figure \ref{fig:beforeandafter1} and Figure \ref{fig:beforeandafter2}, which compare the sample size function (plotted at the boundary of the inequality, i.e. the smallest possible sample size fulfilling Equation (\ref{eq:ss_with_borrow_Wink})) with respect to $w_q$ before and after the functional transformation of $w_q \to f(w_q)=w_q'$. These examples are for the simplest cases of borrowing from one and two historical datasets.
Sample sizes corresponding to $n(w_1=0,w_2=0), n(w_1=1,w_2=0), n(w_1=0,w_2=1)$ and $n(w_1=1,w_1=1)$ remain fixed after the transformation of $w_q \to w'_q$ as required. This will be the case for borrowing from any number of sources, i.e., borrowing from $Q$ sources will have $2^{Q}$ fixed points corresponding to each unique combination of $w_1=\{0, 1\}, ..., w_Q=\{0, 1\}$. Between these fixed points, via the proposed transformation, the change in the sample size is evenly distributed across $w_1, ..., w_Q  \, \epsilon \,  [0, 1]$. As before, the sample size is minimized with full incorporation of information from all sources, i.e., $\{w_1,..., w_Q \}= 0$ and maximized with full discounting of information from both sources, i.e., $\{w_1, ..., w_Q \}= 1$.

\section{Application to the Design of a Randomized Controlled Trial in Alzheimer's Disease}
\label{s:motivatingexample}
In this section, we consider how the proposed method could be applied to determine an appropriate sample size for a hypothetical new trial using real data from several relevant historical RCTs.

Alzheimer's disease (AD) is a chronic age-related illness characterized by cognitive decline. It is the most common form of dementia, with incidence increasing globally due to increasing life expectancy. There are limited pharmaceutical interventions which are effective in reducing symptoms of cognitive decline, however a systematic review by \citet{du2018physical} highlighted that several previous studies have suggested that exercise may slow the progression of cognitive decline in patients with AD. 

Consider planning a new two-arm RCT to investigate whether physical activity can improve cognition in patients with Alzheimer's disease. The two treatments to be compared in the new trial are denoted $T$ (physical activity) and $C$ (standard/usual care). The primary outcome is the difference in treatment group means at a single post-randomization followup timepoint in the Mini Mental State Examination (MMSE) score \citep{arevalo2021mini}. MMSE is a 30-point questionnaire that provides a summary measure of cognitive function where a higher score represents better cognitive performance. It is used extensively in clinical research settings to estimate the severity of impairment, and to document change in impairment over time. Suppose in the new trial that the MMSE of each subject at 4 months post-randomization will be denoted by $y_{ij}, i=1,...,n_j, j = T, C$, and $y_{i,j}$ will be treated as normally distributed with mean $\mu_{j}$ and common (known) variance $\sigma_0^2$, as in Section \ref{Bayes_SS}. The observed difference in means $\Bar{Y}_T-\Bar{Y}_C = \Bar{Y}_\Delta$ is assumed to be normally distributed, $\Bar{Y}_\Delta \sim \mathcal{N} (\mu_{\Delta}, \sigma_0^2/nR(1-R))$ with positive values indicating an advantage for the physical activity group. Based on a recent study of MMSE scores in those with cognitive impairments, $\sigma_0^2 = 3.69^2$ \citep{salis2023mini}.

Consider first a frequentist formulation of the sample size calculation. Suppose we wish to detect a minimum clinically important difference (MCID) between treatment groups of $\delta = 1$ point on the MMSE (it was reported by \citet{mishra2023minimal} that MCID thresholds for MMSE in AD trials are commonly between 1 to 3 points). For a one-sided type I error rate $\alpha = 0.05$ and power $1-\beta=0.80$, the total sample size required is minimized by equal allocation to treatment and control groups, i.e., $R = 0.5$. For these parameters, Equation (\ref{eq:ss_freq}) yields a total sample size of $n=338$ (rounded up to the nearest even integer). Our Bayesian sample size calculation with no borrowing, Equation (\ref{eq:ss_no_borrow}), gives the same result setting a large $s_0^2$ (e.g., $s_0^2 = 100$), with $\eta=0.95$ and $\zeta = 0.80$. 

For obvious reasons, recruiting large numbers of patients onto AD trials might be challenging, with limitations due to ethical and practical issues. Furthermore, high costs can be a concern with trial participants necessarily needing more intense monitoring compared to cognitively intact individuals \citep{chandra2021ethical}.

Now, suppose that data from 7 historical trials is available with which to form an informative prior for $\mu_{\Delta}$, summarized in Table \ref{table:mot_Ex_summary}.
\begin{singlespace}
\begin{table}[H]
\caption{Results of seven historical RCTs measuring MMSE outcomes for individuals with AD, adapted from \citet{du2018physical}. Treatment effects have been summarized in the form of $\lambda_q| \bm{y_q} \sim \mathcal{N} (\theta_q, \tau_q^2)$.}
\label{table:mot_Ex_summary}
\begin{center}
\begin{tabular}{cllcrlcrcc}
\toprule
 & & \multicolumn{2}{c}{Experimental} & & \multicolumn{2}{c}{Control} & &\multicolumn{2}{c}{Difference}\\
 $q$ & Study & Mean  & SD & $n_T$  & Mean & SD & $n_C$ & $\theta_q$ & $\tau_q^2$ \\
  \midrule
  1 & \citet{Vreugdenhil2012} & 23.9 & 5 & 20 & 19 & 7.7 & 20 & 4.90 & 4.21 \\
  2 & \citet{hoffmann2016moderate} & 23.9 & 3.4 & 107 & 23.9 & 3.9 & 93 & 0 & 0.27 \\
  3 & \citet{venturelli2011six} & 12 & 2 & 11 & 6 & 2 & 10 &  6 & 0.76 \\
  4 & \citet{miu2008randomised} & 17.4 & 5.7 & 24 & 19.2 & 4.2 & 28 & -1.8 & 1.89 \\
  5 & \citet{yang2015effects} & 22.83 & 2.75 & 25 & 19.54 & 3.43 & 25 & 3.29 & 0.77 \\
  6 & \citet{holthoff2015effects} & 22.11 & 0.57 & 15 & 20.72 & 0.55 & 15  &1.39  & 0.04 \\
  7 & \citet{kwak2007effect} & 19.1 & 6.5 & 15 & 12.3 & 6.7 & 15 & 6.8 & 5.81 \\
\bottomrule 
\end{tabular}
\end{center}
\end{table} 
\end{singlespace}
It is clear from Table \ref{table:mot_Ex_summary} that there is substantial heterogeneity between studies, therefore with the help of a clinical expert we suppose we have elicited probabilities $w_1, ..., w_Q \, \epsilon \, [0, 1]$ which quantify the irrelevance of each historical trial in respect of the new study.

We note that it might be easier to elicit these quantities as a degree of relevance (rather than degree of skepticism); for example, if an expert thinks data source $q$ is $25\%$ relevant to the new trial then we set $w_q=0.75$. We also note that the proposed methodology assumes that a single expert is consulted, or that multiple experts can agree on single values for $\bm{w_q}$. The process of eliciting and reconciling  multiple expert opinions on probabilities is a complex topic outside the scope of this paper; see, for example, \citet{Hora2016} for an in depth discussion. For illustrative purposes, let us assume that we have elicited a set of probabilities $\bm{w_q} \, \epsilon \,  [0,1]$, with $w_1 = 0.65, w_2 = 0.90, w_3 = 0.75, w_4 = 0.75, w_5 = 0.40, w_6 = 0.95, w_7 = 0.50$. This set would imply a desire to incorporate the most amount of information from source $5$ and least from source $6$.

Firstly, using Equation (\ref{eq:simp_wdash_funcq}) to transform $\bm{w_q} \to \bm{w'_q}$ results in $w'_1 = 7.66 \times 10^{-3}, w'_2 = 2.37 \times 10^{-3}, w'_3 = 2.26 \times 10^{-3}, w'_4 = 5.58 \times 10^{-3}, w'_5 = 5.10 \times 10^{-4}, w'_6 = 7.86 \times 10^{-4}, w'_7 = 5.69 \times 10^{-3}$. By the method in Section \ref{CPP_W}, using $\bm{w'_q}$, $\bm{\theta_q}$ and $\bm{\tau^2_q}$ leads to an informative prior for the treatment effect in the new trial, $\mu_{\Delta}| \bm{y_1},...,\bm{y_7} \sim \mathcal{N} (\theta_{CP^*}, \sigma^2_{CP^*})$, where $\theta_{CP^*} = 2.33$ and $\sigma^2_{CP^*} = 0.34$.  Equation (\ref{eq:ss_with_borrow_Wink}) (setting $a_{01}=1.01, b_{01}=1.01, a_{02}=1 \times 10^6, b_{02}=1$), gives the total sample size (for $\eta = 0.95$ and $\zeta = 0.80$) as $n=176$ (rounded up to the nearest even integer).

Note, that if we just used the `raw' $\bm{w_q}$ in Equation (\ref{eq:ss_with_borrow_Wink}), we would be faced with the issue of over-discounting (described in previous sections), resulting in a sample size of $332$. Also note that if we wished to include information to the specified degree (without transforming $\bm{w_q}$), then we would have to have elicited values of $w_1 = 7.66 \times 10^{-3}, w_2 = 2.37 \times 10^{-3}, w_3 = 2.26 \times 10^{-3}, w_4 = 5.58 \times 10^{-3}, w_5 = 5.10 \times 10^{-4}, w_6 = 7.86 \times 10^{-4}, w_7 = 5.69 \times 10^{-3}$, which would have been very difficult to elicit even if the expert(s) had substantial statistical knowledge. 
\section{Performance Evaluation}
\label{s:performanceevaluation}
We present a brief simulation study where the purpose is to verify that the proposed sample size function and linearization technique achieve the pre-specified statistical properties across a range of scenarios. To be clear, according to the criteria of the Bayesian decision framework in Section \ref{Bayesian_DM}, the sample size should be large enough to guarantee a conclusion of efficacy, such that $\mathbb{P}(\theta > 0 ) \geq \eta$, or, if not, futility, such that $\mathbb{P}(\theta \leq \delta ) \geq \zeta$. The goal of the simulation study therefore is not to compare our sample size formula and linearization technique against another method, but rather to test the hypothesis that by the proposed method $100 \%$ of trials will reach a definitive conclusion.
\subsection{Basic Settings}
\label{basicsettings}
Four contrasting configurations, $A, ..., D$, of hypothetical historical data are investigated, with each containing historical information from 5 independent sources shown in Table \ref{table:scenarios}. We suppose that probabilities $\bm{w_q} = w_1,...,w_5$ have been elicited to implement the proposed approach for borrowing information from each respective source. We fix $\bm{w_q}$ to be the same across all four configurations to facilitate easier comparisons: $w_1 = 0.2, w_2 =0.4, w_3 =0.8, w_4 = 0.6, w_5 = 0.7$. We suppose for demonstration purposes that data from historical data source $1$ is considered particularly relevant to the new trial, setting $w_1 = 0.2$ (in a real situation, this might be for example because the historic trial has been performed most recently, or that it was undertaken at an earlier stage in the same pharmaceutical development pipeline). Sources 2-5 have been considered less relevant with $w_2, ..., w_5$ set accordingly.
\begin{singlespace}\begin{table}[H]
\caption{Configurations $A,..., D$, of hypothetical historical data where mean treatment effect parameter from source $q$ is assumed to have been independently summarized by $\lambda_q \sim \mathcal{N} (\theta_q, \tau_q^2)$. Each source is accompanied by a $w_q$ for borrowing of information, summarizing pre-experimental information about $\mu_{\Delta}$.}
\label{table:scenarios}
\begin{center}
\begin{tabular}{cccp{1.1cm}p{1.1cm}p{1.1cm}p{1.1cm}p{1.1cm}} 
\toprule
   & Config. & & \multicolumn{5}{c}{Historical data source, $q$:} \\
Config. & description & Parameter & 1 & 2 & 3 & 4 & 5 \\
\midrule
   & Weak & $\theta_q$ & 0.10 & 0.24 & 0.37 & 0 & -0.05  \\
 A & historical& $\tau_q^2$ & 1.25 & 0.73 & 0.92 & 1.29 & 0.66 \\ 
  & info. & $w_q$ & 0.20 & 0.40 & 0.80 & 0.60 & 0.70 \\
\midrule
   & & $\theta_q$ & 0 & -0.05 & 2.14 & 0.37 & 1.10  \\
  B & Mixed 1 & $\tau_q^2$ & 1.29 & 0.66 & 0.50 & 0.92 & 0.75 \\
  & & $w_q$ & 0.20 & 0.40 & 0.80 & 0.60 & 0.70 \\
  \midrule
   & & $\theta_q$ & 1.10 & 0.37 & -0.05 & 2.14 & 0 \\
 C & Mixed 2 & $\tau_q^2$ & 0.75 & 0.92 & 0.66 & 0.50 & 1.29\\
  & & $w_q$ & 0.20 & 0.40 & 0.80 & 0.60 & 0.70 \\
 \midrule
   & Strong & $\theta_q$ & 1.10 & 2.14 & 1.07 & 0.60 & 0.85 \\
 D & historical & $\tau_q^2$ & 0.75 & 0.50 & 0.82 & 0.89& 0.26\\
  & info. & $w_q$ & 0.20 & 0.40 & 0.80 & 0.60 & 0.70 \\
\bottomrule
\end{tabular}
\end{center}
\end{table} 
\end{singlespace}
Configuration descriptions classify the nature of the treatment effects observed in the historical trials, with `weak historical info.' meaning low/neutral relative treatment effects observed historically with relatively high variances, and `strong historical info.' indicating more positive historic treatment effects with comparatively smaller variances. Mixed 1 and Mixed 2 use a combination of $\theta_q$ and $\tau_q^2$ from A and D. Weights in Mixed 1 favor the neutral trials, while weights in Mixed 2 favor the more positive trials. Based on the example in Section \ref{s:motivatingexample}, the MCID between treatment arms in the new trial is set to be $\delta = 1$ and we assume a common (known) variance in outcome measures of $\sigma^2_0 = 3.69^2$. Probability boundaries for decision making in terms of efficacy are $\eta=0.95$, and for futility, $\zeta=0.80$. For each configuration of historical data, a sample size is calculated for the new trial: first, using Equation (\ref{eq:simp_wdash_funcq}) to transform $\bm{w_q} \to \bm{w'_q}$, and then Equation (\ref{eq:ss_with_borrow_Wink}), with $\bm{w'_{q}}$ and $\bm{\tau^2_{q}}$ (setting $a_{01}=1.01, b_{01}=1.01, a_{02}=1 \times 10^6, b_{02}=1$). Note that, although $\eta$ and $\zeta$ have been set to be equivalent to the often used $(1-\alpha)$ and $(1 - \beta)$ in the frequentist paradigm (which are set to control type I error rate and power respectively), it must be remembered that these values do not represent the same quantity. As discussed in \citet{Whiteheadetal2008}, there is no reason to assume any form of equivalence since their meanings are fundamentally different. 

For the new trial, we set equal allocation to treatment and control, $R=0.5$.  Outcomes in the control group are generated for each configuration according to $Y_{iC} \sim \mathcal{N} (0, \sigma^2_0), i=1,...n_{k}/2$ (where $k = 1,..., 4$ indexes configurations $A,...,D$). Outcomes in the treatment group are generated according to  $Y_{iT} \sim \mathcal{N} (\mu_{\Delta}, \sigma^2_0), i=1,...n_{k}/2$. 
For each simulation replicate, true treatment effects are set to be one of the following:
\begin{enumerate}
\item Treatment efficacy, $\mu_{\Delta} = 1$. 
\item Treatment futility, $\mu_{\Delta} = 0$.  
\end{enumerate}
A Bayesian analysis model is applied to each simulation replicate, with prior set according to the CP from each configuration. Evidence of treatment efficacy will be concluded if $\mathbb{P}(\mu_{\Delta} > 0) \geq 0.95$. If $\mathbb{P}(\mu_{\Delta} > 0) < 0.95$ then, according to our pre-specified criteria, it should be the case that $\mathbb{P}(\mu_{\Delta} \leq \delta) \geq 0.80$. Results are summarized for $\mu_{\Delta} = 1$ and $\mu_{\Delta} = 0$ respectively by calculating the percentage of trials in which a decisive conclusion can be reached by averaging across 10,000 simulated trial replicates. This results in a total of 8 scenarios. The Bayesian analysis model is fitted analytically using Equations (\ref{eq:d_thetaCPPnew}) and (\ref{eq:precthetaCPPnew}) in R version 4.2.1 (2022-06-23). 
\subsection{Results}
\label{results}
Table \ref{table:res_trans_vals} gives transformed values of $w_q \to w'_q$ via the method described in Section \ref{w_trans_process}.
\begin{singlespace}\begin{table}[H]
\caption{Transformed values of $w_q \to w'_q$.}
\label{table:res_trans_vals}
\begin{center}
\begin{tabular}{ccp{1.9cm}p{1.9cm}p{1.9cm}p{1.9cm}p{1.9cm}} 
\toprule
Config. & $w_q$ & 0.20 & 0.40 & 0.80 & 0.60 & 0.70 \\
  \midrule
 A & $w'_q$ & $3.05 \times 10^{-3}$ & $4.76 \times 10^{-3} $& $3.48 \times 10^{-2}$ & $1.86 \times 10^{-2}$ & $1.49 \times 10^{-2}$\\
 B &  $w'_q$ & $ 3.14\times 10^{-3}$ & $ 4.31\times 10^{-3}$ & $ 1.93\times 10^{-2}$ & $ 1.34\times 10^{-2}$ & $ 1.69\times 10^{-2}$\\
 C & $w'_q$  &  $ 1.84\times 10^{-3}$ & $ 5.98\times 10^{-3}$ & $ 2.53\times 10^{-2}$ & $ 7.33\times 10^{-3}$ & $ 2.86\times 10^{-2}$\\
 D &  $w'_q$  & $1.84 \times 10^{-3}$ & $ 3.27\times 10^{-3}$ & $ 3.12\times 10^{-2}$ & $ 1.29\times 10^{-2}$ & $ 5.96\times 10^{-3}$\\
\bottomrule
\end{tabular}
\end{center}
\end{table}
\end{singlespace}
Table \ref{table:scenarios_res} displays sample sizes (rounded up to the nearest even integer to allow for equal allocation between treatment groups) for each configuration calculated using Equation (\ref{eq:ss_with_borrow_Wink}) with $\bm{w'_q}$ and $\bm{\tau^2_q}$, along with the corresponding prior parameters used for design and analysis. The prior for configuration A is centered closer to zero with a higher variance than the priors for other configurations, resulting in a sample size of $n \geq 204$. The prior for configuration D is the most `enthusiastic', centered on a positive treatment effect with a lower variance, resulting in $n \geq 112$. Configuration B results in a prior centered on a low treatment effect, whereas the prior derived from configuration C is centered on a positive treatment effect. Configurations B and C result in priors with similar variances.
\begin{singlespace}
\begin{table}[H]
\caption{Priors for treatment effect in the new experiment, $\mu_{\Delta} \sim \mathcal{N}(\theta_{CP^*}, \sigma^2_{CP^*})$, along with corresponding sample sizes (rounded up to nearest even integer for $R=0.5$) for configurations $A,..., D$.}
\label{table:scenarios_res}
\begin{center}
\begin{tabular}{cccc}
\toprule
 Config.  & $\theta_{CP^*}(\bm{w'_{q}}, \bm{\theta_{q}}, \bm{\tau^2_{q}})$ & $\sigma^2_{CP^*}(\bm{w'_{q}}, \bm{\tau^2_{q}})$ &  $n$ \\
\midrule
  A &  0.131 & 0.405 & 204 \\
  B & 0.515 & 0.358 & 186 \\
  C & 1.015 & 0.325 & 170 \\
  D & 1.276 & 0.242 & 112 \\
\bottomrule
\end{tabular}
\end{center}
\end{table}
\end{singlespace}
 Table \ref{table:percentage_eff_vs_fut} displays the percentage of simulated trials concluding that experimental treatment is efficacious (\% Eff.) or futile (\% Fut.) for each configuration in scenarios where $\mu_{\Delta} =1$ and $\mu_{\Delta} =0$ respectively.  The percentage efficacious is defined as the percentage of trials out of 10,000 simulations in which $\mathbb{P}(\mu_{\Delta}>0)\geq 0.95$, while the percentage futile is the percentage of trials out of 10,000 simulations in which $\mathbb{P}(\mu_{\Delta}>0)<0.95$ and $\mathbb{P}(\mu_{\Delta}\leq 0.4) \geq 0.80$. The total percentage is $100\%$ in all scenarios, demonstrating that the pre-specified statistical properties are upheld by the proposed method.
\begin{singlespace}\begin{table}[H]
\caption{Percentage of simulated trials that conclude treatment is efficacious or futile when $\mu_{\Delta}=1$ and $\mu_{\Delta}=0$ (analyzed using informative priors for $\mu_{\Delta}$ as specified in Table \ref{table:scenarios_res}.)}
\label{table:percentage_eff_vs_fut}
\begin{center}
\begin{tabular}{cccccccc}
\toprule
 & & \multicolumn{2}{c}{$\mu_{\Delta}=1$:} &  & \multicolumn{2}{c}{$\mu_{\Delta}=0$:} & \\
 Config. & $n$ & \% Eff.  & \% Fut. & Total \%  & \% Eff. & \% Fut. & Total \% \\
  \midrule
  A & 204 & 49.3 & 50.7 & 100 & 2.6 & 97.4 & 100 \\
  B & 186 & 66.0 & 34.0 & 100 & 7.3 & 92.7 & 100 \\
  C & 170 & 88.7 & 11.3 & 100 & 29.2 & 70.8 & 100 \\
  D & 112 & 98.7 & 1.3 & 100 & 79.8 & 20.2 & 100 \\
\bottomrule
\end{tabular}
\end{center}
\end{table}
\end{singlespace}
We emphasize that an investigation of frequentist operating characteristics was not the purpose of this section. Nonetheless, as anticipated, and as mentioned in Section \ref{s:intro}, it is clear from Table \ref{table:percentage_eff_vs_fut} that to realize the benefits of historical borrowing (at least, in traditional frequentist terms), the treatment effect in the new trial should be similar to the treatment effect in historical trials. When this is the case, we observe higher `power' (as in configurations C and D when $\mu_{\Delta} = 1$) (or a lower `type I error rate', as in configurations A and B when $\mu_{\Delta} = 0$) by borrowing of information. However, this necessarily comes at the risk of a higher type I error rate / reduced power when there is a high degree of heterogeneity between historical and current trials. As discussed in \citet{Kopp-Schneider2020}, if one wishes to control the type I error rate in the traditional sense, all prior information must be disregarded in the analysis. In any practical application therefore, careful selection of historical trials for inclusion as well as extensive simulations at the trial design stage would be necessary. 
\section{Discussion}
\label{s:discussion}
Our central goal in this paper has been twofold: firstly, to offer a solution for the problem of nonmonotonic behavior of discrepancy weights caused by the prior aggregation method proposed in \citet{Zheng2023a, Zheng2023b}. Our proposed alternative ensures that discrepancy weights behave monotonically with respect to the amount of information included from a particular source. This leads us to derive a Bayesian sample size formula and achieve our second goal of linearization to improve interpretability. Following our methodology, given a set of historical data sources, clinical expert(s) only have to specify probabilities representing the (dis)similarity between each historic data source and the current trial ($\bm{w_q}$) for a trial statistician to then to incorporate the specified amount of information (using $\bm{w'_q}$). We hope that these ideas can encourage effective communication between statisticians and subject-matter experts to elicit sensible values for these weights.

There are a number of ways in which this work could be extended/generalized. One possibility would be extension to other Bayesian methods proposed for clinical trials which utilize weights for borrowing, such as the robust MAP prior \citep{SchmidliRobustMAP2014}.  More broadly, the methodology could be applied in any research area (not just clinical trials) where it would be desirable to design an experiment using information from previous studies or external data. 

We note that in this work we have assumed independence of historical data sources as a simplified case of aggregating information by the method of \citet{winkler1981combining}, in which a method is proposed in the case that historical sources are dependent. When historical studies are conducted on distinct patients the independence assumption would seem reasonable. However, if the historical data relate to multiple trials in the same patients (for example, phase II/III trials), then the dependence between studies could easily be accounted for by the same method in \citet{winkler1981combining}, with calculation of the pairwise correlations between sources.

Focus in this work is on the design of a two-armed trial where there is prior information on the difference in means between treatment and control arms. We acknowledge that it is more common for methods for borrowing from historical data to consider borrowing only on the control arm (i.e., using  historical control information to augment or replace a concurrent control). The methods presented here could be adapted to this case such that a prior would be formed for the arm-based statistic(s). As noted in  \citet{Zheng2023a}, selection of historical data on a single arm should be done carefully to avoid bias that may affect the inference of the difference in means. 
Our sample size formula and linearization technique could also be extended to other clinical trial designs where borrowing can be incorporated; for example, combined phase II/III trials using borrowing from the phase II part of the trial to reduce the sample size for the phase III part, or a basket trial setting (for concurrent borrowing between subtrials) in which a sample size is sought for each subtrial, $k = 1,..., K$, with sample sizes being solved as a system of $K$ simultaneous equations.

The proposed methodology utilizes a single prior for both the design and analysis of the new experiment. There may be instances where it is desirable to modify the analysis prior according to the observed similarity between the historic datasets and current trial. In this case, a distributional distance metric $\epsilon \, [0,1]$ such as the Hellinger distance \citep{dey1994robust} might be useful in updating $\bm{w_q}$ for the analysis. However, as noted in \citet{Zheng2023b}, this would affect the properties of the Bayesian decision making framework on which the sample size formula is based. Specifically, when $\bm{w_q}$ are set to larger values in the analysis than in the design (i.e., less borrowing is implemented than planned), it may not be possible to reach a decisive conclusion regarding efficacy or futility. Conversely, using smaller $\bm{w_q}$ in the analysis than the design (i.e., more borrowing is implemented than planned) would lead to a more precise posterior distribution which may have a higher risk of bias.

In our approach we have restricted focus to known variance in outcome measure, $\sigma_0^2$ (common in many settings), and we approximated $\nu^{-1}_q$ by making some simplifying assumptions, which resulted in a closed form for the sample size calculation. One avenue for development would be a more fully Bayesian approach in which priors are specified for $\sigma_0$ and/or $\nu_q$. Furthermore, in this paper we have focused on the Bayesian decision making framework proposed in \citet{Whiteheadetal2008}, however, it would be simple to adapt the sample size formula for consideration of other Bayesian properties. For example, a sample size formula controlling average properties of posterior interval probabilities could be achieved in a similar manner as in \citet{Zheng2023a}, where a sample size formula is proposed for control of the average coverage criterion (ACC) or the average length criterion (ALC); for implementation of our method this would simply require replacing the prior precision ($\sigma^{-2}_{CP}$) proposed in  \citet{Zheng2023a} with our alternative proposal ($\sigma^{-2}_{CP^*}$).

In conclusion, historical data from a range of sources are often available in the planning of a new trial, but inclusion of such data for study design and analysis is not common practice. Part of the reason might be difficulty in interpretability of discrepancy parameters. We hope our work will help to bridge this gap and encourage uptake of these innovative methods, however we caution that consideration of sample size on its own should not be the only factor when determining whether a borrowing method is appropriate. Simulation is generally still needed to evaluate its performance (bias, power, type I error, etc.).

\vspace*{-8pt}
\begin{singlespace}
\section*{Software}
R code for reproducing the Motivating Example and Performance Evaluation is posted online at GitHub: \url{https://github.com/lou-e-whitehead/BayesianSSD_2024}.

\section*{Acknowledgements}
Dr Zheng's contribution to this work was supported by Cancer Research UK (RCCPDF\textbackslash100008). James M. S. Wason is funded by NIHR Research Professorship (NIHR301614). \vspace*{-8pt}

\section*{Supplementary Materials}
\label{s:suppmat}
\subsection*{A. Bayesian sample size formula for a two-armed RCT with binary endpoints}
\label{binaryRCT_ext}
Consider planning a new RCT to compare response rates between a treatment and control group using relevant historical data. We follow similar steps detailed in the Supplementary Material in \citet{Zheng2023b}, with two important differences: 
\begin{itemize}
    \item The sample size calculation here is for a two-arm RCT with borrowing from historical data, rather than a basket trial with multiple substudies borrowing from concurrent data.
    \item Our new formulation of $\sigma^{-2}_{CP^*}$, i.e.,
    \begin{align}
    \sigma^{-2}_{CP^*} &= \sum_{q=1}^{Q} \xi^{-2}_{q} \nonumber \\
    &=  \sum_{q=1}^{Q} \left[\left(\tau_q^2 + \frac{w_{q} b_{01}}{a_{01}-1}+\frac{(1-w_{q}) b_{02}}{a_{02}-1}\right)^{-1}\right].
\end{align}
    detailed in Section 3.1 of the main paper. 
\end{itemize} 

In the new trial, the observed proportion of responders in each arm of the sample is
\begin{equation*}
    \hat{\rho}_{j}=\sum^{n_{j}}_{i=1} Y_{ij} / n_{j}
\end{equation*}
where $Y_{ij}$ is a binary indicator denoting response ($=1$) or non-response ($=0$) for subject $i = 1,...,n_{j}$ in arm $j = T, C$. By the central limit theorem, and as long as the proportions are not close to $0$ or $1$, the log odds ratio of the sample
\begin{equation*}
   \log{(\hat{OR})} = \log{\left(\frac{\hat{\rho}_{T} (1-\hat{\rho}_{C})}{\hat{\rho}_{C} (1-\hat{\rho}_{T})} \right)}
\end{equation*}
would be approximately normally distributed \citep{agresti2003categorical},
\begin{equation*}
    \log{(\hat{OR})} 
    \, \dot\sim \, \mathcal{N} \left( \log \left(\frac{\rho_{T}(1-\rho_{C})}{\rho_{C}(1-\rho_{T})} \right), \frac{1}{n_T} \left( \frac{1}{\rho_{T}} + \frac{1}{1 - \rho_{T}}\right) +\frac{1}{n_C} \left( \frac{1}{\rho_{C}} + \frac{1}{1 - \rho_{C}}\right) \right)
\end{equation*}
We further let $\mu_{\Delta}  = \log \left(\frac{\rho_{T}(1-\rho_{C})}{\rho_{C}(1-\rho_{T})} \right)$ and simplify the variance of $\log{(\hat{OR})} $,
\begin{equation*}
    \frac{1}{n_T} \left( \frac{1}{\rho_{T}} + \frac{1}{1 - \rho_{T}}\right) +\frac{1}{n_C} \left( \frac{1}{\rho_{C}} + \frac{1}{1 - \rho_{C}}\right) = \frac{1}{n} \left(\frac{1}{\rho_{T}(1-\rho_T)} + \frac{1}{\rho_{C}(1-\rho_C)} \right)
\end{equation*}
where we have assumed that $n_T=n_C=n$. 
Following the methodology proposed in the main paper, we aggregate historical data from sources $q=1,...,Q$ which has been summarized in the form $\lambda_q  
    \, \dot\sim \, \mathcal{N} (\theta_q, \tau_q^2)$ (where $\lambda_q$ now represent log-odds ratios) into a single prior for $\mu_{\Delta}$, 
\begin{equation*}
    \mu_{\Delta}| \bm{y_1},...,\bm{y_Q} \, \dot\sim \, \mathcal{N} (\theta_{CP^*},  \sigma^2_{CP^*}),
\end{equation*}
where, as in the main paper,
\begin{equation*}
    \theta_{CP^*} = \sum_{q=1}^{Q} p^{*}_{q}\theta_q, \ \
    \sigma^2_{CP^*} = \left(\sum_{q=1}^{Q} \xi^{-2}_{q}\right)^{-1},
\end{equation*}
\begin{equation*}
    p^{*}_q = \frac{\xi^{-2}_q} {\left(\sum_{q=1}^{Q} \xi^{-2}_{q}\right)},
\end{equation*}
\begin{equation*}
    \xi_q^{-2} = \left[ \tau_q^2 + \frac{w_{q} b_{01}}{a_{01}-1}+\frac{(1-w_{q}) b_{02}}{a_{02}-1} \right]^{-1}.
\end{equation*}
The prior is updated to a posterior with the new trial data,
\begin{equation*}
    \mu_{\Delta}| \bm{y_1},...,\bm{y_Q}, \bm{y_{new}} \, \dot\sim \, \mathcal{N}\left(d_{\theta^*}, \sigma^2_{\theta^*} \right),
\end{equation*}
where  
\begin{equation*}
    d_{\theta^*} = \frac{\sigma^{-2}_{CP^*} \cdot \theta_{CP^*} + \log{(\hat{OR})} \cdot n \left(\frac{1}{\rho_{T}(1-\rho_T)} + \frac{1}{\rho_{C}(1-\rho_C)} \right) ^{-1} } {\sigma^{-2}_{CP^*}+n \left(\frac{1}{\rho_{T}(1-\rho_T)} + \frac{1}{\rho_{C}(1-\rho_C)} \right) ^{-1} }
\end{equation*}
and  
\begin{equation*}
 \sigma^2_{\theta^*}=\left(\sigma^{-2}_{CP^*} + n \left(\frac{1}{\rho_{T}(1-\rho_T)} + \frac{1}{\rho_{C}(1-\rho_C)} \right) ^{-1} \right)^{-1}.   
\end{equation*}
Applying the same decision criterion with a clinically meaningful effect size $\delta$, the minimum sample size needed for (each arm in) the new study can be found as:
\begin{equation*}
    n \geq \left(\frac{1}{\rho_{T}(1-\rho_T)} + \frac{1}{\rho_{C}(1-\rho_C)} \right)  \left(  \left(\frac{z_{\eta}+z_{\zeta}}{\delta} \right)^2 - \sigma^{-2}_{CP^*} \right). 
\end{equation*}
\subsection*{B. Bayesian sample size formula for a two-armed RCT with time-to-event endpoints}
\label{TTERCT_ext}
We now consider extending the proposed sample size formula for planning an RCT with a time-to-event outcome incorporating borrowing from historical data. As in the previous example, the derivation follows similar steps as the Supplementary Material in \citet{Zheng2023b} with the same key differences already noted. For simplicity, we follow \citet{george1974planning} to assume that event times from the new trial $T_{ij}$ are exponentially distributed:
\begin{equation*}
    T_{ij} \sim \exp{(\pi_{j})}\,\,\,
    i=1,...,n_q; \,\, j=E, C
\end{equation*}
with rate $\pi_{j} > 0$ (and the treatment arm now coded `$E$' for clarity). The average time-to-event observed in treatment group $j$  is $\bar{T}_{j}$ and the corresponding number of events is denoted $D_{j}$. By the central limit theorem,
\begin{equation*}
    \bar{T}_{j} \, \dot\sim \, \mathcal{N} \left( \frac{1}{\pi_{j}}, \frac{1}{D_{j}\pi_{j}} \right).
\end{equation*}
By the delta method,
\begin{equation*}
    \log{(\bar{T}_{j})} \, \dot\sim \, \mathcal{N} \left( -\log{(\pi_{j})}, \frac{1}{D_{j}} \right),
\end{equation*}
therefore,
\begin{equation*}
    \log{\left(\frac{\bar{T}_{E}}{\bar{T}_{C}}\right)} \, \dot\sim \, \mathcal{N} \left( \log{\left(\frac{\pi_{E}}{\pi_{C}}\right)}, \frac{1}{D_{E}} +\frac{1}{D_{C}} \right).
\end{equation*}
We further let $\mu_{\Delta}  =  \log{\left(\frac{\pi_{E}}{\pi_{C}}\right)}$. Following the methodology proposed in the main paper, we aggregate the historical data into a single prior for $\mu_{\Delta}| \bm{y_1},...,\bm{y_Q} \, \dot\sim \, \mathcal{N} (\theta_{CP^*},  \sigma^2_{CP^*})$. It is assumed that historical data from sources $q=1,...,Q$ has been summarized in the form $\lambda_q  
    \, \dot\sim \, \mathcal{N} (\theta_q, \tau_q^2)$ (where $\lambda_q$ now represent log ratios of mean event times),
\begin{equation*}
    \theta_{CP^*} = \sum_{q=1}^{Q} p^{*}_{q} \theta_q, \ \
    \sigma^2_{CP^*} = \left(\sum_{q=1}^{Q} \xi^{-2}_{q}\right)^{-1},
\end{equation*}
\begin{equation*}
    p^{*}_q = \frac{\xi^{-2}_q} {\left(\sum_{q=1}^{Q} \xi^{-2}_{q}\right)},
\end{equation*}
\begin{equation*}
    \xi_q^{-2} = \left[ \tau_q^2 + \frac{w_{q} b_{01}}{a_{01}-1}+\frac{(1-w_{q}) b_{02}}{a_{02}-1} \right]^{-1}.
\end{equation*}
The prior is updated to a posterior with the new trial data,
\begin{equation*}
    \mu_{\Delta}| \bm{y_1},...,\bm{y_Q}, \bm{y_{new}} \, \dot\sim \, \mathcal{N}\left(d_{\theta^*}, \sigma^2_{\theta^*} \right),
\end{equation*}
where
\begin{equation*}
    d_{\theta^*} = \frac{\sigma^{-2}_{CP^*} \cdot \theta_{CP^*} +  \log{\left(\frac{\bar{T}_{E}}{\bar{T}_{C}}\right)} \cdot (\frac{1}{D_{E}} +\frac{1}{D_{C}}) ^{-1} } {\sigma^{-2}_{CP^*}+(\frac{1}{D_{E}} +\frac{1}{D_{C}}) ^{-1}}
\end{equation*}
and  
\begin{equation*}
 \sigma^2_{\theta^*}=\left(\sigma^{-2}_{CP^*} + \left(\frac{1}{D_{E}} +\frac{1}{D_{C}} \right) ^{-1} \right)^{-1}.   
\end{equation*}
Applying the same decision criterion as the main paper, the minimum sample size required in the new trial can be found according to
\begin{equation*}
    \frac{D_E D_C}{D_E + D_C} \geq \left(\frac{z_{\eta}+z_{\zeta}}{\delta} \right)^2 - \sigma^{-2}_{CP^*}.
\end{equation*}
Equivalently,
\begin{equation*}
    D \geq \frac{1}{R(1-R)}\left[\left(\frac{z_{\eta}+z_{\zeta}}{\delta} \right)^2 - \sigma^{-2}_{CP^*}\right],
\end{equation*}
where $\delta$  is a target clinically meaningful effect size, $D=D_E+D_C$ denotes the total number of events required and $R$ is the proportion of the sample randomized to the treatment arm $E$. We note that further work would be needed to derive a sample size formula for time-to-event data with censoring.

\subsection*{C. Bayesian sample size formula for a single-arm trial with a binary outcome}
\label{sing_bin_ext}
In early phase oncology trials single-arm designs with binary outcomes are frequently conducted. Here we consider planning such a trial incorporating borrowing from historical data. Suppose in the new trial all patients are given a new treatment and observed to be either a `responder’ or `non-responder’ according to predetermined criteria. Letting $Y$ denote the number of responders and $n$ the number of patients needed in the new trial. This leads to $\mathbb{E}(Y)=np$ and $Var(Y)=np(1-p)$ where $p$ is the proportion of responders and $p \, \epsilon \, [0,1]$. By the delta method, the estimator of log-odds is asymptotically normally distributed,
\begin{equation*}
    \log{\left(\frac{\hat{p}}{1-\hat{p}}\right)} \, \dot\sim \, \mathcal{N}\left(\log{\left(\frac{{p}}{1-{p}}\right)}, \, np(1-p) \right)
\end{equation*}
We further let $\mu_{\Delta} = \log{\left(\frac{{p}}{1-{p}}\right)}$. Following the methodology proposed in the main paper, we aggregate the historical data relating to the same arm into a single prior for $\mu_{\Delta}| \bm{y_1},...,\bm{y_Q} \, \dot\sim \, \mathcal{N} (\theta_{CP^*},  \sigma^2_{CP^*})$, where
\begin{equation*}
    \theta_{CP^*} = \sum_{q=1}^{Q} p^{*}_{q}\log{\left(\frac{\hat{p}_q}{1-\hat{p}_q}\right)} , \ \
    \sigma^2_{CP^*} = \left(\sum_{q=1}^{Q} \xi^{-2}_{q}\right)^{-1},
\end{equation*}
\begin{equation*}
    p^{*}_q = \frac{\xi^{-2}_q} {\left(\sum_{q=1}^{Q} \xi^{-2}_{q}\right)},
\end{equation*}
\begin{equation*}
    \xi_q^{-2} = \left[ \tau_q^2 + \frac{w_{q} b_{01}}{a_{01}-1}+\frac{(1-w_{q}) b_{02}}{a_{02}-1} \right]^{-1}.
\end{equation*}
The prior is updated to a posterior with the new trial data,
\begin{equation*}
    \mu_{\Delta}| \bm{y_1},...,\bm{y_Q}, \bm{y_{new}} \, \dot\sim \, \mathcal{N}\left(d_{\theta^*}, \sigma^2_{\theta^*} \right),
\end{equation*}
where
\begin{equation*}
    d_{\theta^*} = \frac{\sigma^{-2}_{CP^*} \cdot \theta_{CP^*} +  \log{\left(\frac{\hat{p}}{1-\hat{p}}\right)}  \cdot (np(1-p)) ^{-1} } {\sigma^{-2}_{CP^*}+(np(1-p)) ^{-1}}
\end{equation*}
and  
\begin{equation*}
 \sigma^2_{\theta^*}=\left(\sigma^{-2}_{CP^*} + (np(1-p)) ^{-1} \right)^{-1}.   
\end{equation*}
Applying the same decision criterion with a clinically meaningful effect size $\delta$, the minimum sample size needed for the new study can be found as:
\begin{equation*}
    n \geq \frac{1}{p(1-p)}\left[ \left(\frac{z_{\eta}+z_{\zeta}}{\delta} \right)^2 - \sigma^{-2}_{CP^*} \right]. 
\end{equation*}

\bibliography{Bibliography}

\providecommand{\noopsort}[1]{}
\begin{thebibliography}{}

\bibitem[Agresti, 2003]{agresti2003categorical}
Agresti, A. (2003).
\newblock {\em Categorical data analysis}.
\newblock Wiley Series in Probability and Statistics. Wiley.

\bibitem[Arevalo-Rodriguez et~al., 2021]{arevalo2021mini}
Arevalo-Rodriguez, I., Smailagic, N., Roqu{\'e}-Figuls, M., Ciapponi, A., Sanchez-Perez, E., Giannakou, A., Pedraza, O.~L., Cosp, X.~B., and Cullum, S. (2021).
\newblock Mini-mental state examination ({MMSE}) for the early detection of dementia in people with mild cognitive impairment ({MCI}).
\newblock {\em Cochrane Database of Systematic Reviews}, 7(7).

\bibitem[Bornkamp, 2012]{Bornkamp2012}
Bornkamp, B. (2012).
\newblock Functional uniform priors for nonlinear modeling.
\newblock {\em Biometrics}, 68(3):893--901.

\bibitem[Bornkamp, 2014]{Bornkamp2014}
Bornkamp, B. (2014).
\newblock Practical considerations for using functional uniform prior distributions for dose-response estimation in clinical trials.
\newblock {\em Biometrical Journal}, 56(6):947--962.

\bibitem[Chandra et~al., 2021]{chandra2021ethical}
Chandra, M., Harbishettar, V., Sawhney, H., and Amanullah, S. (2021).
\newblock Ethical issues in dementia research.
\newblock {\em Indian Journal of Psychological Medicine}, 43(5\_suppl):S25--S30.

\bibitem[Dey and Birmiwal, 1994]{dey1994robust}
Dey, D.~K. and Birmiwal, L.~R. (1994).
\newblock Robust {B}ayesian analysis using divergence measures.
\newblock {\em Statistics \& Probability Letters}, 20(4):287--294.

\bibitem[Dias et~al., 2017]{dias2017elicitation}
Dias, L.~C., Morton, A., and Quigley, J. (2017).
\newblock {\em Elicitation: The Science and Art of Structuring Judgement}.
\newblock International Series in Operations Research \& Management Science. Springer International Publishing.

\bibitem[Dky et~al., 2008]{miu2008randomised}
Dky, M., Szeto, S., Mak, Y., et~al. (2008).
\newblock A randomised controlled trial on the effect of exercise on physical, cognitive and affective function in dementia subjects.
\newblock {\em Asian J Gerontol Geriatr}, 3(1):8--16.

\bibitem[Du et~al., 2018]{du2018physical}
Du, Z., Li, Y., Li, J., Zhou, C., Li, F., and Yang, X. (2018).
\newblock Physical activity can improve cognition in patients with {A}lzheimer’s disease: a systematic review and meta-analysis of randomized controlled trials.
\newblock {\em Clinical Interventions in Aging}, pages 1593--1603.

\bibitem[George and Desu, 1974]{george1974planning}
George, S.~L. and Desu, M. (1974).
\newblock Planning the size and duration of a clinical trial studying the time to some critical event.
\newblock {\em Journal of chronic diseases}, 27(1-2):15--24.

\bibitem[Grinstead and Snell, 1997]{grinstead1997introduction}
Grinstead, C.~M. and Snell, J.~L. (1997).
\newblock {\em Introduction to probability}.
\newblock American Mathematical Soc.

\bibitem[Hampson et~al., 2014]{Hampson2014}
Hampson, L.~V., Whitehead, J., Eleftheriou, D., and Brogan, P. (2014).
\newblock Bayesian methods for the design and interpretation of clinical trials in very rare diseases.
\newblock {\em Statistics in Medicine}, 33(24):4186--4201.

\bibitem[Hariton and Locascio, 2018]{Hariton2018}
Hariton, E. and Locascio, J.~J. (2018).
\newblock Randomised controlled trials – the gold standard for effectiveness research.
\newblock {\em BJOG: An International Journal of Obstetrics \& Gynaecology}, 125(13):1716--1716.

\bibitem[Hobbs et~al., 2011]{hobbs2011hierarchical}
Hobbs, B.~P., Carlin, B.~P., Mandrekar, S.~J., and Sargent, D.~J. (2011).
\newblock Hierarchical commensurate and power prior models for adaptive incorporation of historical information in clinical trials.
\newblock {\em Biometrics}, 67(3):1047--1056.

\bibitem[Hobbs et~al., 2012]{Hobbs2012CommensPrior}
Hobbs, B.~P., Sargent, D.~J., and Carlin, B.~P. (2012).
\newblock {Commensurate Priors for Incorporating Historical Information in Clinical Trials Using General and Generalized Linear Models}.
\newblock {\em Bayesian Analysis}, 7(3):639 -- 674.

\bibitem[Hoffmann et~al., 2016]{hoffmann2016moderate}
Hoffmann, K., Sobol, N.~A., Frederiksen, K.~S., Beyer, N., Vogel, A., Vestergaard, K., Br{\ae}ndgaard, H., Gottrup, H., Lolk, A., Wermuth, L., et~al. (2016).
\newblock Moderate-to-high intensity physical exercise in patients with {A}lzheimer’s disease: a randomized controlled trial.
\newblock {\em Journal of Alzheimer's Disease}, 50(2):443--453.

\bibitem[Holthoff et~al., 2015]{holthoff2015effects}
Holthoff, V.~A., Marschner, K., Scharf, M., Steding, J., Meyer, S., Koch, R., and Donix, M. (2015).
\newblock Effects of physical activity training in patients with {A}lzheimer’s dementia: results of a pilot {RCT} study.
\newblock {\em PloS one}, 10(4):e0121478.

\bibitem[Hora, 2016]{Hora2016}
Hora, S. (2016).
\newblock {497. Probability Elicitation}.
\newblock In {\em {The Oxford Handbook of Probability and Philosophy}}. Oxford University Press.

\bibitem[Ibrahim and Chen, 2000]{ibrahim2000power}
Ibrahim, J.~G. and Chen, M.-H. (2000).
\newblock Power prior distributions for regression models.
\newblock {\em Statistical Science}, pages 46--60.

\bibitem[Johnson et~al., 2010]{johnson2010methods}
Johnson, S.~R., Tomlinson, G.~A., Hawker, G.~A., Granton, J.~T., and Feldman, B.~M. (2010).
\newblock Methods to elicit beliefs for {B}ayesian priors: a systematic review.
\newblock {\em Journal of clinical epidemiology}, 63(4):355--369.

\bibitem[Julious, 2023]{julious2023sample}
Julious, S.~A. (2023).
\newblock {\em Sample sizes for clinical trials}.
\newblock CRC Press.

\bibitem[Kopp-Schneider et~al., 2020]{Kopp-Schneider2020}
Kopp-Schneider, A., Calderazzo, S., and Wiesenfarth, M. (2020).
\newblock Power gains by using external information in clinical trials are typically not possible when requiring strict type i error control.
\newblock {\em Biometrical Journal}, 62(2):361--374.

\bibitem[Kwak et~al., 2007]{kwak2007effect}
Kwak, Y.-S., Um, S.-Y., Son, T.-G., and Kim, D.-J. (2007).
\newblock Effect of regular exercise on senile dementia patients.
\newblock {\em International Journal of Sports Medicine}, pages 471--474.

\bibitem[Mishra et~al., 2023]{mishra2023minimal}
Mishra, B., Sudheer, P., Agarwal, A., Srivastava, M. V.~P., Vishnu, V.~Y., et~al. (2023).
\newblock Minimal clinically important difference ({MCID}) in patient-reported outcome measures for neurological conditions: Review of concept and methods.
\newblock {\em Annals of Indian Academy of Neurology}.

\bibitem[Neuenschwander et~al., 2010]{NeuenschwanderMAP2010}
Neuenschwander, B., Capkun-Niggli, G., Branson, M., and Spiegelhalter, D.~J. (2010).
\newblock Summarizing historical information on controls in clinical trials.
\newblock {\em Clinical Trials}, 7(1):5--18.
\newblock PMID: 20156954.

\bibitem[Pocock, 1976]{POCOCK1976175}
Pocock, S.~J. (1976).
\newblock The combination of randomized and historical controls in clinical trials.
\newblock {\em Journal of Chronic Diseases}, 29(3):175--188.

\bibitem[Salis et~al., 2023]{salis2023mini}
Salis, F., Costaggiu, D., and Mandas, A. (2023).
\newblock Mini-mental state examination: optimal cut-off levels for mild and severe cognitive impairment.
\newblock {\em Geriatrics}, 8(1):12.

\bibitem[Schmidli et~al., 2014]{SchmidliRobustMAP2014}
Schmidli, H., Gsteiger, S., Roychoudhury, S., O'Hagan, A., Spiegelhalter, D., and Neuenschwander, B. (2014).
\newblock Robust meta-analytic-predictive priors in clinical trials with historical control information.
\newblock {\em Biometrics}, 70(4):1023--1032.

\bibitem[Venturelli et~al., 2011]{venturelli2011six}
Venturelli, M., Scarsini, R., and Schena, F. (2011).
\newblock Six-month walking program changes cognitive and {ADL} performance in patients with {A}lzheimer's.
\newblock {\em American Journal of Alzheimer's Disease \& Other Dementias{\textregistered}}, 26(5):381--388.

\bibitem[Viele et~al., 2014]{Viele2014}
Viele, K., Berry, S., Neuenschwander, B., Amzal, B., Chen, F., Enas, N., Hobbs, B., Ibrahim, J.~G., Kinnersley, N., Lindborg, S., Micallef, S., Roychoudhury, S., and Thompson, L. (2014).
\newblock Use of historical control data for assessing treatment effects in clinical trials.
\newblock {\em Pharmaceutical Statistics}, 13(1):41--54.

\bibitem[Vreugdenhil et~al., 2012]{Vreugdenhil2012}
Vreugdenhil, A., Cannell, J., Davies, A., and Razay, G. (2012).
\newblock A community-based exercise programme to improve functional ability in people with {A}lzheimer’s disease: a randomized controlled trial.
\newblock {\em Scandinavian Journal of Caring Sciences}, 26(1):12--19.

\bibitem[Wadsworth et~al., 2018]{Wadsworth2018}
Wadsworth, I., Hampson, L.~V., and Jaki, T. (2018).
\newblock Extrapolation of efficacy and other data to support the development of new medicines for children: A systematic review of methods.
\newblock {\em Statistical Methods in Medical Research}, 27(2):398--413.
\newblock PMID: 26994211.

\bibitem[Whitehead et~al., 2008]{Whiteheadetal2008}
Whitehead, J., Valdés-Márquez, E., Johnson, P., and Graham, G. (2008).
\newblock Bayesian sample size for exploratory clinical trials incorporating historical data.
\newblock {\em Statistics in Medicine}, 27(13):2307--2327.

\bibitem[Winkler, 1981]{winkler1981combining}
Winkler, R.~L. (1981).
\newblock Combining probability distributions from dependent information sources.
\newblock {\em Management Science}, 27(4):479--488.

\bibitem[Yang et~al., 2015]{yang2015effects}
Yang, S.-Y., Shan, C.-L., Qing, H., Wang, W., Zhu, Y., Yin, M.-M., Machado, S., Yuan, T.-F., and Wu, T. (2015).
\newblock The effects of aerobic exercise on cognitive function of {A}lzheimer’s disease patients.
\newblock {\em CNS \& Neurological Disorders-Drug Targets (Formerly Current Drug Targets-CNS \& Neurological Disorders)}, 14(10):1292--1297.

\bibitem[Zheng et~al., 2023a]{Zheng2023a}
Zheng, H., \noopsort{A}Jaki, T., and Wason, J. M.~S. (2023a).
\newblock Bayesian sample size determination using commensurate priors to leverage preexperimental data.
\newblock {\em Biometrics}, 79(2):669–683.

\bibitem[Zheng et~al., 2023b]{Zheng2023b}
Zheng, H., \noopsort{B}Grayling, M.~J., Mozgunov, P., Jaki, T., and Wason, J. M.~S. (2023b).
\newblock Bayesian sample size determination in basket trials borrowing information between subsets.
\newblock {\em Biostatistics}, 24(4):1000 -- 1016.

\bibitem[Zheng and Wason, 2022]{ZhengandWason2022}
Zheng, H. and Wason, J. M.~S. (2022).
\newblock {Borrowing of information across patient subgroups in a basket trial based on distributional discrepancy}.
\newblock {\em Biostatistics}, 23(1):120--135.

\end{thebibliography}
\end{singlespace}
\end{document}